\documentclass[trackchanges, twocolumn]{aastex701}

\usepackage{underscore}
\usepackage{CJK}

\begin{document}

\title{Detection of CO, H$_2$O, and OH in WASP-18b with JWST/NIRISS using Direct-Extracted Spectra and Cross-Correlation}

\author[orcid=0009-0001-1490-2991, sname='Ouyang']{Qinglin Ouyang}
\affiliation{National Key Laboratory of Deep Space Exploration / Department of Astronomy, University of Science and Technology of China, Hefei 230026, People’s Republic of China}
\email{qlouyang@ustc.edu.cn}  

\author[orcid=0000-0001-9585-9034, sname='Yan']{Fei Yan} 
\affiliation{National Key Laboratory of Deep Space Exploration / Department of Astronomy, University of Science and Technology of China, Hefei 230026, People’s Republic of China}
\email[show]{yanfei@ustc.edu.cn}

\author[orcid=0009-0006-7386-3634]{Shuo Liu}
\affiliation{National Key Laboratory of Deep Space Exploration / Department of Astronomy, University of Science and Technology of China, Hefei 230026, People’s Republic of China}
\email{sureliu@mail.ustc.edu.cn}

\author[orcid=0009-0009-5056-6639]{Boyue Guo}
\affiliation{National Key Laboratory of Deep Space Exploration / Department of Astronomy, University of Science and Technology of China, Hefei 230026, People’s Republic of China}
\email{gby@mail.ustc.edu.cn}

\author[orcid=0000-0003-0740-5433]{Guo Chen}
\affiliation{CAS Key Laboratory of Planetary Sciences, Purple Mountain Observatory, Chinese Academy of Sciences, Nanjing 210023, People’s Republic of China}
\email{guochen@pmo.ac.cn}

\author[orcid=0000-0003-0987-1593]{Enric Pall\'{e}}
\affiliation{Instituto de Astrofísica de Canarias, E-38200 La Laguna, Tenerife, Spain}
\affiliation{Departamento de Astrofísica, Universidad de La Laguna, 38206 La Laguna, Tenerife, Spain}
\email{epalle@iac.es}

\author[orcid=0009-0001-5682-5015]{Yuanheng Yang}
\affiliation{CAS Key Laboratory of Planetary Sciences, Purple Mountain Observatory, Chinese Academy of Sciences, Nanjing 210023, People’s Republic of China}
\affiliation{School of Astronomy and Space Science, University of Science and Technology of China, Hefei 230026, People’s Republic of China}
\email{yhyang@pmo.ac.cn}

\author[orcid=0000-0002-9702-4441]{Wei Wang}
\affiliation{CAS Key Laboratory of Optical Astronomy, National Astronomical Observatories, Chinese Academy of Sciences, Datun Road A20, Beijing 100101, China}
\email{wangw@nao.cas.cn}

\author[orcid=0000-0003-1207-3787]{Meng Zhai}
\affiliation{Chinese Academy of Sciences South America Center for Astronomy (CASSACA), National Astronomical Observatories, CAS, Datun Road A20, Beijing 100101, China}
\email{mzhai@nao.cas.cn}

\author[orcid=0009-0003-4854-3610]{Qian Chen}
\affiliation{CAS Key Laboratory of Optical Astronomy, National Astronomical Observatories, Chinese Academy of Sciences, Datun Road A20, Beijing 100101, China}
\affiliation{School of Astronomy and Space Science, University of Chinese Academy of Sciences (UCAS), Yuquan Road A19, Beijing 100049, China}
\email{chenqian@nao.cas.cn}

\begin{abstract}

The James Webb Space Telescope (JWST) has revolutionized the characterization of exoplanetary atmospheres, offering unprecedented sensitivity to probe their chemical and physical properties. Recently, a growing trend has emerged to obtain atmospheric information directly from pixel-level planetary spectra. In this work, we re-analyzed the WASP-18b NIRISS/SOSS dataset by employing a direct extraction method. This new method preserves the spectral information at the native instrumental resolution, thereby enabling the application of cross-correlation techniques and providing atmospheric retrievals with enhanced precision and richer information content. With this methodology, we report detections of CO at $4.4\sigma$ significance, H$_2$O at $3.4\sigma$, and OH at $7.8\sigma$, where CO and OH were previously unseen. Building on these unambiguous detections, our subsequent retrieval analysis significantly improves the constraints on atmospheric abundances. Our results demonstrate that the cross-correlation technique effectively extracts molecular signals from medium-resolution JWST data, enhancing detection sensitivity. By revisiting JWST archival data with cross-correlation and retrieval analysis, we can achieve a more comprehensive survey of planetary atmospheric chemistry, thereby placing precise constraints on key parameters such as planetary metallicity and C/O ratio.

%as well as the horizontal and vertical thermal structure. This absence has limited the ability to place precise constraints on the planet's carbon-to-oxygen (C/O) ratio. 

%with \textbf{direct spectrum-extraction and cross-correlation techniques}. The new method successfully yields reliable signals of CO at $4.4\sigma$ significance, H$_2$O at $3.4\sigma$, and OH at $7.8\sigma$, \textbf{where CO and OH were previously unseen}. Building on these unambiguous detections, our subsequent retrieval analysis significantly improves the constraints on atmospheric abundances, leading to a better constraint on the C/O ratio for WASP-18b. Our results demonstrate that the cross-correlation technique effectively extracts molecular signals from medium-resolution JWST data, enhancing detection sensitivity. \textbf{By revisiting JWST archival data with cross-correlation and retrieval analysis, we can achieve a more comprehensive survey of planetary atmospheric chemistry, thereby placing precise constraints on key parameters such as planetary metallicity and C/O ratio.} %This work not only maximizes the scientific return of the JWST mission but also can establish a new methodological foundation for theories on planetary population diversity and formation mechanisms.

\end{abstract}



\section{Introduction} \label{sec:Intro}

The James Webb Space Telescope (JWST) is fundamentally reshaping our understanding of exoplanetary atmospheres \citep{Espinoza2025}. Its unprecedented sensitivity and wavelength coverage allow for the detection of diverse atmospheric species, including H$_2$O \citep{Taylor2023, Ahrer2023}, CO$_2$ \citep{Alderson2023}, CO \citep{Rustamkulov2023}, CH$_4$ \citep{Benneke2024}, SO$_2$ \citep{Tsai2023, Gressier2025}, SiO \citep{Gapp2025}, VO, and H$^-$ \citep{Pelletier2026}. Beyond species detection, these observations enable the inference of cloud properties \citep{Grant2023, Inglis2024}, photochemistry \citep{Tsai2023}, and disequilibrium chemical processes \citep{Evans-Soma2025}, as well as the characterization of atmospheric escape \citep{Krishnamurthy2025}. 
%Moreover, JWST data are providing novel insights into the internal structures of these worlds, particularly for sub-Neptunes \citep{Sing2024, Holmberg2024, Hu2025}.

The standard procedure for characterizing transiting exoplanet atmospheres with JWST typically involves light curve fitting and atmospheric retrieval. Light curve fitting extracts the planetary spectrum from the time-series stellar spectra, while atmospheric retrieval matches planetary spectrum against theoretical models. During the planetary spectrum extraction, to secure a sufficient signal-to-noise ratio (S/N) for spectroscopic light curve fitting to robustly extract the planetary signal (transit or eclipse depth), stellar spectra are often binned over dozens of pixels. This binning process degrades the spectral resolution, potentially diluting high-frequency signals from specific molecular species. Furthermore, in retrieval analysis, the space of molecular combinations available for model construction is vast and highly degenerate. Only a limited subset of these combinations is typically explored for a given detection \citep{Welbanks2025}. Consequently, the identification of specific species often relies on statistical model comparison, yet the detection significance derived from the resulting Bayes factors often represents an overestimated upper limit \citep{Kipping2025, Thorngren2025}.

Cross-correlation techniques offer a robust solution to these challenges. Recent studies have successfully adapted cross-correlation techniques \citep{Snellen2010} to JWST NIRSpec G395H and G235H datasets, confirming detections of CO, H$_2$O, and CO$_2$ in the transmission spectrum of WASP-39b\citep{Esparza-Borges2023, Esparza-Borges2025}, and molecular carbon (C$_3$ and C$_2$) in PSR J2322–2650b’s emission spectrum \citep{Zhang2025}. Their work demonstrates that cross-correlation techniques are capable of capturing the high-frequency signals of specific molecules even within JWST medium-resolution ($R \sim 2,700$) spectra. However, the application of such techniques to other JWST instruments, such as the Near Infrared Imager and Slitless Spectrograph (NIRISS; \citealt{NIRISS_intro}), has yet to be explored in detail.

In this work, we test the performance of cross-correlation techniques on the NIRISS single object slitless spectroscopy (NIRISS/SOSS; \citet{SOSS_intro}) dataset, utilizing secondary eclipse observations of WASP-18b from the JWST Transiting Exoplanet Community Early Release Science Program (JTEC ERS-1366; \citealt{Coulombe2023}). WASP-18b is an ultra-hot Jupiter with $M_{\mathrm{p}} = 10.20 \pm 0.35 M_{\mathrm{Jup}}$, $R_{\mathrm{p}} = 1.240 \pm 0.079 R_{\mathrm{Jup}}$, and an equilibrium temperature of $T_{\mathrm{eq}} \sim 2,400$ K, orbiting a bright ($J_{\mathrm{mag}} = 8.4$) F6V-type host star \citep{Hellier2009, Cortes-Zuleta2020}. \citet{Coulombe2023} reported the dayside thermal emission spectrum of WASP-18b, detecting clear H$_2$O emission features and potential signatures of H$^-$, TiO, and VO. Furthermore, \citet{Challener2025} used the same dataset to perform eclipse mapping, interpreting the horizontal and vertical thermal structure and chemical composition of the atmosphere. However, neither study resolved the CO signal, which has been clearly detected in previous ground-based high-resolution spectroscopic observations \citep{Brogi2023, Yan2023}, resulting in limited constraints on the C/O ratio. Therefore, we aim to verify the detection of CO and other species in the WASP-18b NIRISS dataset using standard cross-correlation techniques, thereby evaluating the capability of this method for NIRISS data and refining the constraints on the planet's atmospheric composition.

%% The "ht!" tells LaTeX to put the figure "here" first, at the "top" next
%% and to override the normal way of calculating a float position.
%% The asterisk after "figure" tells the compiler to span multiple columns
%% if a two column style is selected.

\section{Data reduction and direct spectrum extraction method} \label{sec:obs reduction}

We analyzed the NIRISS/SOSS secondary eclipse observation of WASP-18b from the JTEC ERS-1366 program \citep{Coulombe2023}.
%The secondary eclipse of WASP-18b was observed by NIRISS/SOSS as part of the JTEC ERS-1366 program. The observation monitored the star for a total duration of 6.71 hours, including a 2.83-hour baseline before the eclipse ingress and a 1.70-hour baseline after the egress. The data were obtained using the SUBSTRIP96 subarray mode (96 $\times$ 2,048 pixels), which covers the first spectral order from 0.85 to 2.85 µm. The time series consists of 2,720 continuous integrations, with each integration containing three groups and lasting 8.88 seconds \citep{Coulombe2023}.
The raw dataset was reduced by the \texttt{exoTEDRF} pipeline (formerly \texttt{supreme-SPOON}; \citep{exoTEDRF, Radica2023, Feinstein2023}), following a procedure similar to that described in \citet{Coulombe2023}. We processed the \textit{uncal} files through Stages 1 to 3 of \texttt{exoTEDRF}. In Stage 1, we performed detector-level processing, which included superbias subtraction, saturation detection, zodiacal background subtraction, linearity correction, jump detection, and ramp fitting. \texttt{exoTEDRF} treats 1/f noise at the group level. We employed the scale-achromatic method, constructing difference images by subtracting a median stack from each individual frame and then subtracting the median value from each column. In Stages 2 and 3, we performed further spectroscopic processing, such as flat field division and bad pixel correction. The centroids of all three SOSS orders were located using the edgetrigger algorithm \citep{Radica2022}. Finally, the stellar spectra were extracted using a simple box aperture with a width of 30 pixels. We only utilized the Order 1 spectra for further analysis. For the wavelength solution, \texttt{exoTEDRF} generated an interpolated PHOENIX spectrum \citep{Husser2013} based on the given stellar parameters of WASP-18 ($T_{\mathrm{eff}}=6400$ K, $\log g=4.367$, [M/H]=0.10), and cross-correlated the extracted spectra with it to derive the ``exoTEDRF" wavelength solution.

To verify the accuracy of the wavelength solution provided by \texttt{exoTEDRF}, we compared the 1D stellar spectra extracted by the pipeline with the PHOENIX stellar model spectrum. We identified a significant deviation of several nanometers between the two. Given that our subsequent cross-correlation analysis is highly sensitive to the precision of the wavelength solution, we further optimized the wavelength calibration by matching the absorption lines of the observed stellar spectrum with those of the model (see Appendix \ref{sec:wave solution} and Figure \ref{fig:steller_spec} for details).

\begin{figure*}[ht!]
\plotone{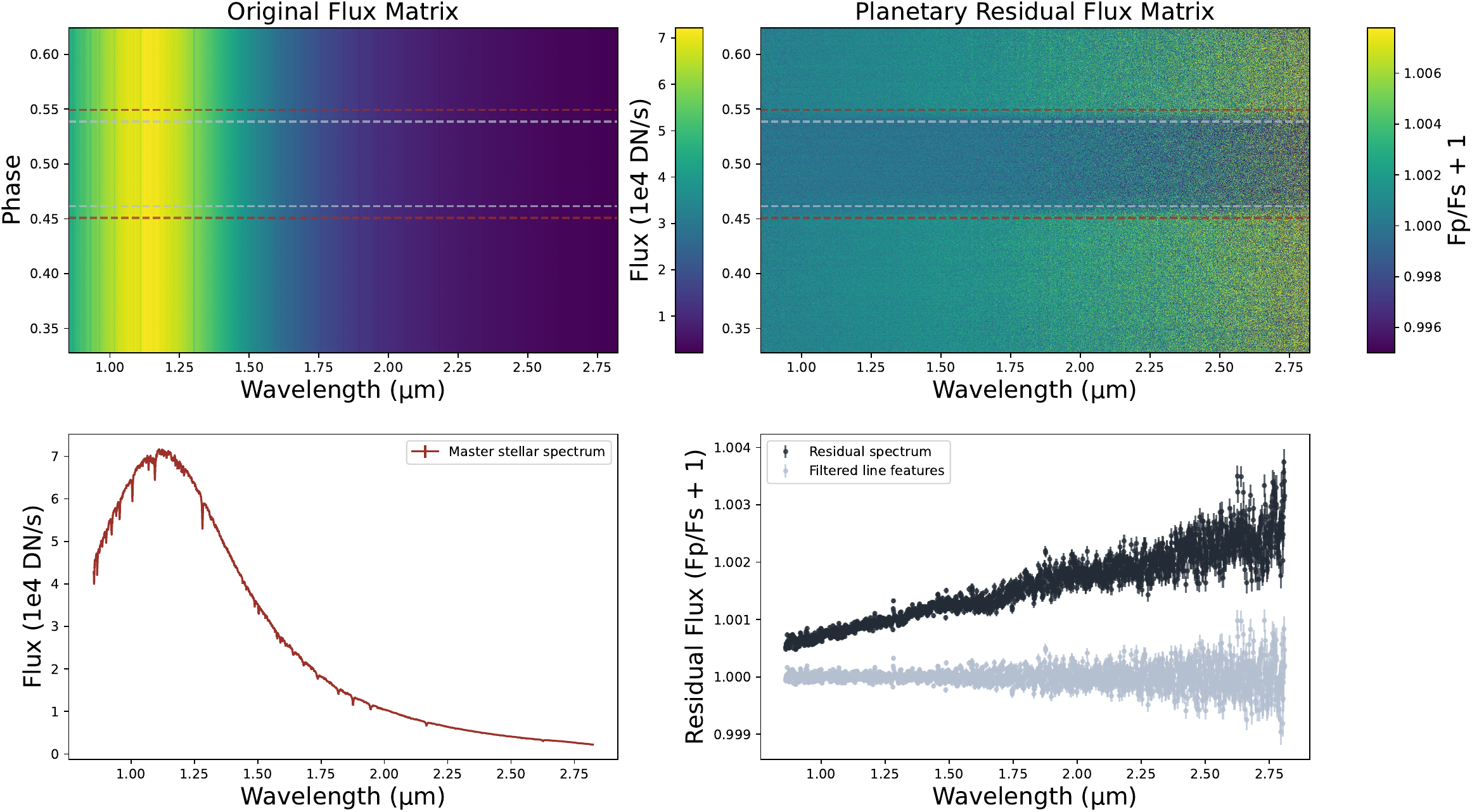}
\caption{The planetary direct-extracted spectrum extraction steps. \textit{Top left}: The original stellar spectral matrix. The red dashed lines indicate $T_1$ and $T_4$, while the white dashed lines mark $T_2$ and $T_3$. \textit{Bottom left}: The master stellar spectrum, derived by averaging all spectra within the eclipse phase ($T_{23}$). \textit{Top right}: The planetary residual flux matrix, obtained by dividing the original flux matrix by the master stellar spectrum. \textit{Bottom right}: The final WASP-18b direct-extracted spectrum (black) and the Gaussian filtered spectrum (light blue).
\label{fig:planet_spec}}
\end{figure*}

To prepare the data for the subsequent cross-correlation analysis, we need to extract a pixel-level thermal emission spectrum of WASP-18b. Our approach differs from the spectroscopic light curve fitting used in \citet{Coulombe2023} and \citet{Esparza-Borges2023}; instead, we employed a direct extraction method conceptually similar to the approaches of \citet{Lustig2025} and Yang et al. (2026, \textit{in prep.}), which parallels the techniques used in high-resolution cross-correlation spectroscopy (HRCCS) to separate planetary and stellar signals \citep{Snellen2010}. First, we arranged all stellar spectra according to their orbital phase. We discarded the 1,884th spectrum, as it was flagged as an anomaly by \texttt{exoTEDRF}'s Principal Component Analysis. We then created a master stellar spectrum by averaging all spectra obtained fully within the secondary eclipse. This master spectrum ($F_\mathrm{s}$), representing the pure stellar signal, was divided out from all spectra in the time series to obtain the planetary residual spectra ($F_\mathrm{p} / F_\mathrm{s} + 1$) (see Fig. \ref{fig:planet_spec}). We note that \citet{Coulombe2023} reported a primary mirror tilt event, which caused a sudden flux drop in their light curves \citep{Rigby2023, Alderson2023}. Our subsequent analysis involves normalizing the extracted spectra, which can effectively mitigate the impact of such systematics on our results.

Additionally, as a result of the wavelength optimization detailed in Appendix \ref{sec:wave solution} (where spectra were aligned to the PHOENIX model), all of our spectra were already in the stellar rest frame. We then corrected these residual spectra for the planet's orbital motion by shifting them according to the expected radial velocity semi-amplitude ($K_p = 236$ km s$^{-1}$) of WASP-18b \citep{Maxted2013, Yan2023}. This step aligned all residual spectra to the planet's rest frame. Finally, we averaged all the out-of-eclipse (not including ingress and egress) phase-aligned and frame-shifted residual spectra to obtain the direct-extracted spectrum (DES), which was used for our subsequent cross-correlation analysis (as shown in Fig. \ref{fig:planet_spec}).

\section{Cross-correlation analysis} \label{sec:CCF}

To perform the cross-correlation analysis with our extracted planetary spectrum, we generated several template emission spectra of WASP-18b using \texttt{petitRADTRANS 3} \citep{Molliere2019, Blain2024}. For the planetary parameters, we set $R_{\mathrm{p}} = 1.240$ $R_{\mathrm{Jup}}$ and $\log g = 4.215$ cm s$^{-2}$ \citep{Cortes-Zuleta2020}. We adopted the median temperature profile (T-P) from the 1D radiative–convective–thermochemical equilibrium grid retrieval in \citet{Coulombe2023}. This profile consists of 44 atmospheric layers, uniformly spaced in log-pressure, spanning from $10^{2.6}$ bar to $10^{-6}$ bar. For the atmospheric composition, we assumed a bulk atmosphere with a mean molecular weight of 2.33, dominated by H$_2$ and He. We also considered the opacity of target species: CO \citep{Rothman2010}, H$_2$O \citep{Polyansky2018}, and OH \citep{Rothman2010}. We utilized the line-by-line radiative transfer mode to generate the templates at an initial resolving power of $R = 10,000$. These high-resolution spectra were then convolved down to the instrumental resolution of NIRISS/SOSS. Considering the resolution of SOSS varies linearly with wavelength (R $\sim 500 - 1,400$), we developed a variable resolution convolution method to convolve the high-resolution spectra. The detailed convolution settings can be found in Appendix \ref{sec:convolution}.
%, Fe\citep{Kurucz2018}, and K\citep{Piskunov1995}

\begin{figure*}[h!]
\plotone{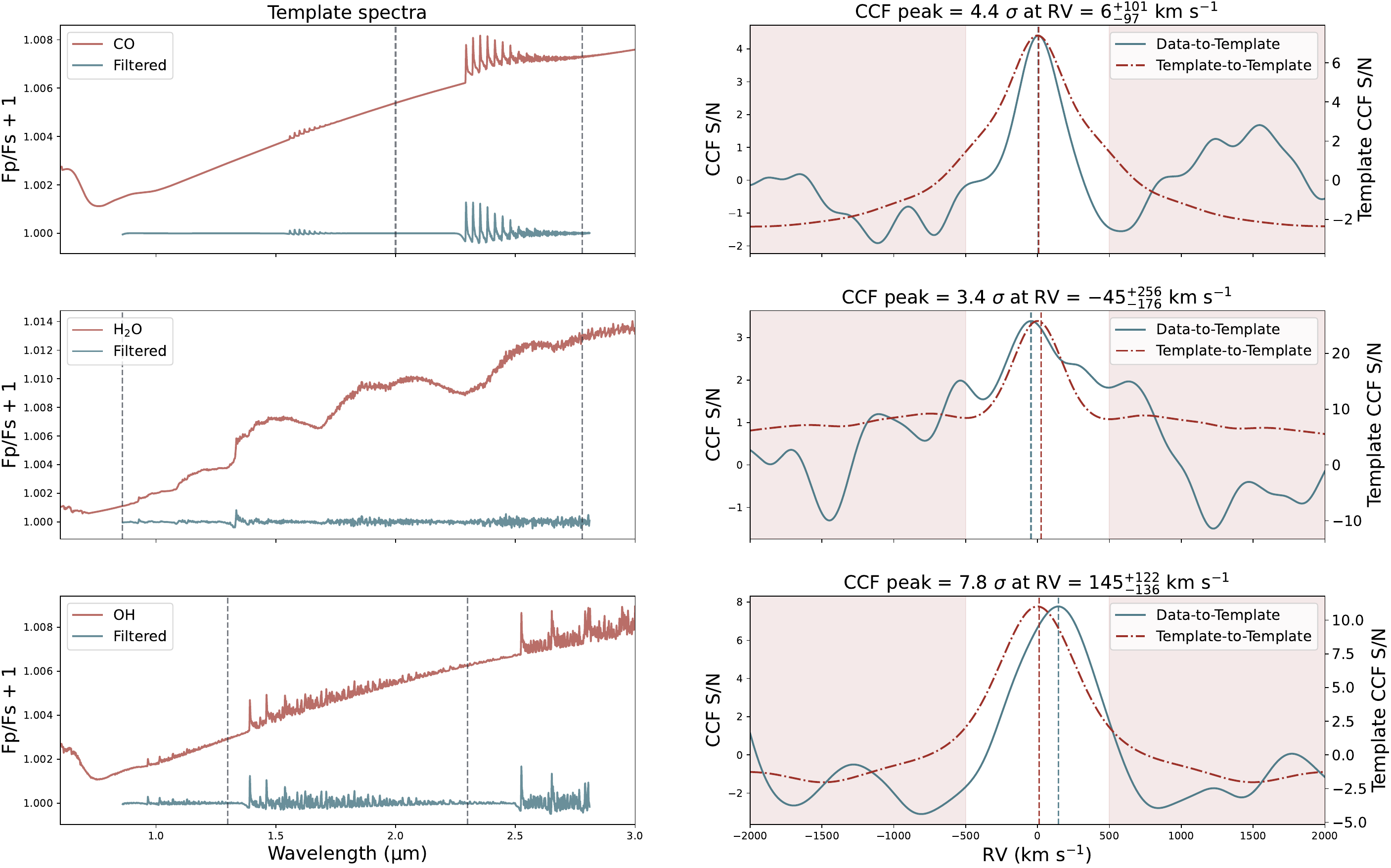}
\caption{Cross-correlation detections from the WASP-18b direct-extracted spectrum. \textit{Left panel}: The template spectra and the Gaussian filtered spectra for CO, H$_2$O, and OH (top to bottom), generated using the methods described in Sec.\ref{sec:CCF}. The vertical dashed lines indicate the wavelength ranges used for the cross-correlation analysis: [2.0–2.78] $\mu$m for CO, [0.86–2.78] $\mu$m for H$_2$O, and [1.30–2.30] $\mu$m for OH. \textit{Right panel}: The data-to-template and template-to-template CCFs S/N for each species, derived from correlating the master planetary residual spectrum with the respective template. The vertical dashed lines mark the radial velocity of the peak signal. The light red shaded areas represent the RV ranges, [–2,000, –500] km s$^{-1}$ and [500, 2,000] km s$^{-1}$, used for noise estimation.
\label{fig:temp_CCF}}
\end{figure*}

We then generated the templates across different radial velocities (RV) based on these models. First, to express the template in the same $F_\mathrm{p}/F_\mathrm{s} + 1$ format as our observed residual spectra, we added the planetary emission template ($F_\mathrm{p}$) to a $T_{\mathrm{eff}}=6400$ K blackbody spectrum (representing $F_\mathrm{s}$). This combined spectrum ($F_\mathrm{p} + F_\mathrm{s}$) was then divided by the blackbody spectrum ($F_\mathrm{s}$) alone. We then constructed the full template grid by shifting these $1 + F_\mathrm{p}/F_\mathrm{s}$ spectra over a RV range from -2,000 km s$^{-1}$ to +2,000 km s$^{-1}$ in 1 km s$^{-1}$ steps. Each shifted template in this grid was then interpolated to the same wavelength points as the observed spectrum. Following this, we applied a Gaussian high-pass filter with a standard deviation of 15 pixels to each template. Finally, we performed the cross-correlation by calculating the standard cross-correlation function (CCF), which for a given radial velocity $v$ is defined as:
\begin{equation}
CCF(v) = \sum_{i} r_{i} \times m_{i}(v)
\end{equation}
where $r_i$ is the value of the high-pass filtered master planetary residual spectrum at wavelength pixel $i$ (processed with the same Gaussian high-pass filter as the template), and $m_i(v)$ is the value of the corresponding filtered template spectrum at the same pixel $i$, which has been Doppler-shifted by velocity $v$. The model template spectra used for this analysis and the corresponding data-to-template CCFs are shown in Figure \ref{fig:temp_CCF}. To mitigate cross-contamination from line blending at the NIRISS resolution, we restricted the templates to specific wavelength ranges where each species' features are relatively isolated.

As mentioned in Sec. \ref{sec:obs reduction}, our direct-extracted spectrum has been shifted to the planet's rest frame. Therefore, if atmospheric species such as CO are present in the atmosphere of WASP-18b, we would expect to detect a CCF peak at an RV near 0 km s$^{-1}$. As shown in Figure \ref{fig:temp_CCF}, both CO and H$_2$O exhibit clear peaks near RV = 0 km s$^{-1}$, while the peak of OH is located at RV = 150 km s$^{-1}$, showing a slight deviation from the expected velocity. We further estimated the noise by using the regions far from the CCF peak (shaded areas in Fig. \ref{fig:temp_CCF}) and converted the CCF values to S/N to assess the significance and uncertainties of the peak RV. This yields detection significances of $4.4\sigma$ at RV $= 10^{+100}_{-100}$ km s$^{-1}$ (for CO), $3.4\sigma$ at RV $= -50^{+260}_{-180}$ km s$^{-1}$ (for H$_2$O) and $7.8\sigma$ at RV = $150^{+120}_{-140}$ km s$^{-1}$ (for OH). We also calculated the template-to-template CCFs, all of which show the expected peak at 0 km s$^{-1}$. Additionally, we investigated the CCF results for TiO and VO, as potential detections of these species were reported by \citet{Coulombe2023}. However, TiO and VO showed no significant signals at the expected RV, likely due to their broad spectral absorption features, making the cross-correlation method less sensitive to them.
%This is likely because these two species primarily exhibit broad absorption band features in the NIRISS/SOSS data, making the cross-correlation method less sensitive to them.

To confirm that these signals originate from the planet, we followed the method of \citet{Cont2022, Yan2022}, aligning the planetary residual spectra to a given $K_p$ value in a grid from -2,000 km s$^{-1}$ to 2,000 km s$^{-1}$, in 5 km s$^{-1}$ steps. Each residual spectrum was then cross-correlated with the model grid, thereby generating the $K_p$ maps for CO, H$_2$O, and OH. To estimate the detection significance, we adopted a method similar to \citet{Brogi2023} by fitting Gaussian distributions to the CCF values and used the standard deviation of the fitted Gaussian as a proxy for the noise. The S/N was then calculated by dividing the CCF values by this noise estimate.

\begin{figure*}[ht!]
\plotone{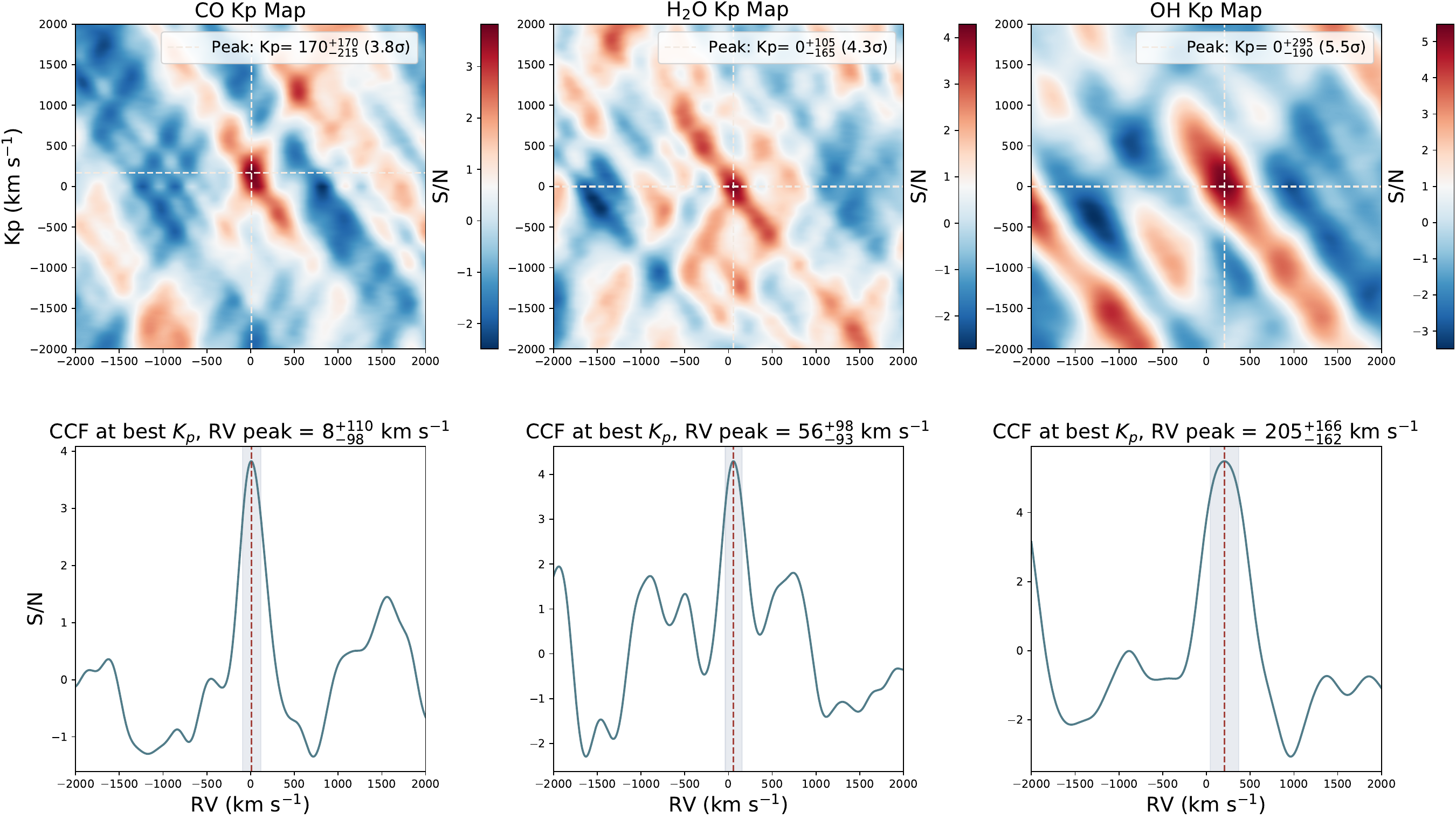}
\caption{\textit{Left panel}: The $K_p$ map for the CO signal of WASP-18b (top) and the corresponding CCF at maximum S/N (bottom). The crossing of the dashed white lines indicates the location of the maximum S/N. In the bottom plot, the vertical dashed line and shaded region mark the RV peak and its 1$\sigma$ confidence interval, respectively. \textit{Middle and Right panels}: Same as the left panel, but for H$_2$O and OH, respectively.
\label{fig:Kp_maps}}
\end{figure*}

The resulting $K_p$ maps are shown in Figure \ref{fig:Kp_maps}. We detect CO, H$_2$O, and OH signals in the dayside atmosphere of WASP-18b at $3.9\sigma$,$ 4.3\sigma$, and $5.5\sigma$, respectively. For the CO signal, the maximum S/N occurs at $K_p = 170^{+170}_{-220}$ km s$^{-1}$ and RV $= 10^{+110}_{-100}$ km s$^{-1}$, which is consistent with previous ground-based high-resolution spectroscopic observations \citep{Yan2023}. The H$_2$O and OH signals, however, peak at $K_p = 0^{+110}_{-170}$ km s$^{-1}$ and $K_p = 0^{+300}_{-190}$ km s$^{-1}$, showing slight deviation from the expected $K_p = 236$ km s$^{-1}$. To rule out the stellar origin for these signals, we cross-correlated the in-eclipse stellar spectrum with the three template spectra. The details can be found in Appendix \ref{sec:stellar_contam}. We did not find significant peaks for CO, H$_2$O, and OH at 0 km s$^{-1}$, confirming that these molecules are absent in the photosphere or potential starspots of WASP-18. We also note that the uncertainties for both $K_p$ and RV are on the order of $\sim 100$ km s$^{-1}$, which corresponds to the radial velocity sampling of a single pixel in JWST/NIRISS spectrum. This indicates that the limited velocity precision of NIRISS/SOSS makes accurate measurement of the true $K_p$ challenging. Nevertheless, our results are robust enough to confirm the planetary origin of these species, demonstrating the feasibility of the cross-correlation method for JWST/NIRISS data and highlighting its reliability for detecting atmospheric species despite the limited velocity precision.

\section{Atmospheric Retrieval} \label{sec:retrieval}

Our analysis in Sec. \ref{sec:CCF} demonstrated that CO, H$_2$O, and OH signals can be robustly detected at the instrumental resolution of JWST/NIRISS. We then applied retrieval analysis to the direct-extracted spectrum to determine the abundances of these atmospheric components and thus place better constraints on the planetary carbon-to-oxygen ratio (C/O) and metallicity ([M/H]).

Before retrieval, it is important to note that our DES is not strictly the entire dayside emission spectrum of WASP-18b. The DES represents the arithmetic mean of all out-of-eclipse planetary residual spectra across different phases. It is subject to geometric effects associated with phase variations and contains systematic variation of the continuum flux. Therefore, the DES has a slight deviation from the emission spectrum derived via light curve fitting (light-curve fitting spectrum, LCS) (see Appendix \ref{sec:brightness} for details). Given that the continuum reliability of the DES is inferior to that of the LCS, using the DES in isolation for retrieval could potentially bias chemical abundance determinations. Therefore, we adopted a ``joint retrieval" methodology. We utilized the LCS from \citet{Coulombe2023} to provide continuum information, while using our filtered DES to provide spectral line profile information. These two datasets were then fit jointly to constrain the atmospheric properties of WASP-18b simultaneously. The log-likelihood function ($\ln {L}$) is expressed as:
\begin{equation}
\ln L_{\text{total}} = \ln L_{\text{cont}} + \ln L_{\text{line}}
\end{equation}
where $\ln L_{\text{cont}}$ is the log-likelihood for the continuum spectrum, and $\ln L_{\text{line}}$ is the log-likelihood for our filtered planetary residual spectrum. Assuming Gaussian-distributed errors, each component is defined by the standard equation:
\begin{equation}
\ln L = -\frac{1}{2} \sum_{i} \left[ \frac{(D_i - M_i)^2}{s_i^2} + \ln(2\pi s_i^2) \right]
\end{equation}
where, for a given dataset, $D_i$ is the data at wavelength $i$, $M_i$ is the corresponding model value, and $s_i$ is the uncertainty. For $\ln L_{\text{cont}}$, $D_i$ and $s_i$ are the flux point and uncertainty of the LCS in \citet{Coulombe2023}, and $M_i$ is the full forward model scaled by the hotspot area fraction $A_{HS}$. For $\ln L_{\text{line}}$, $D_i$ represents the flux point of our filtered DES, $s_i$ is the uncertainty propagated from the original DES flux before filtering, and $M_i$ is the identically filtered forward model scaled by $A_{HS}$.

\begin{figure*}[ht!]
\plotone{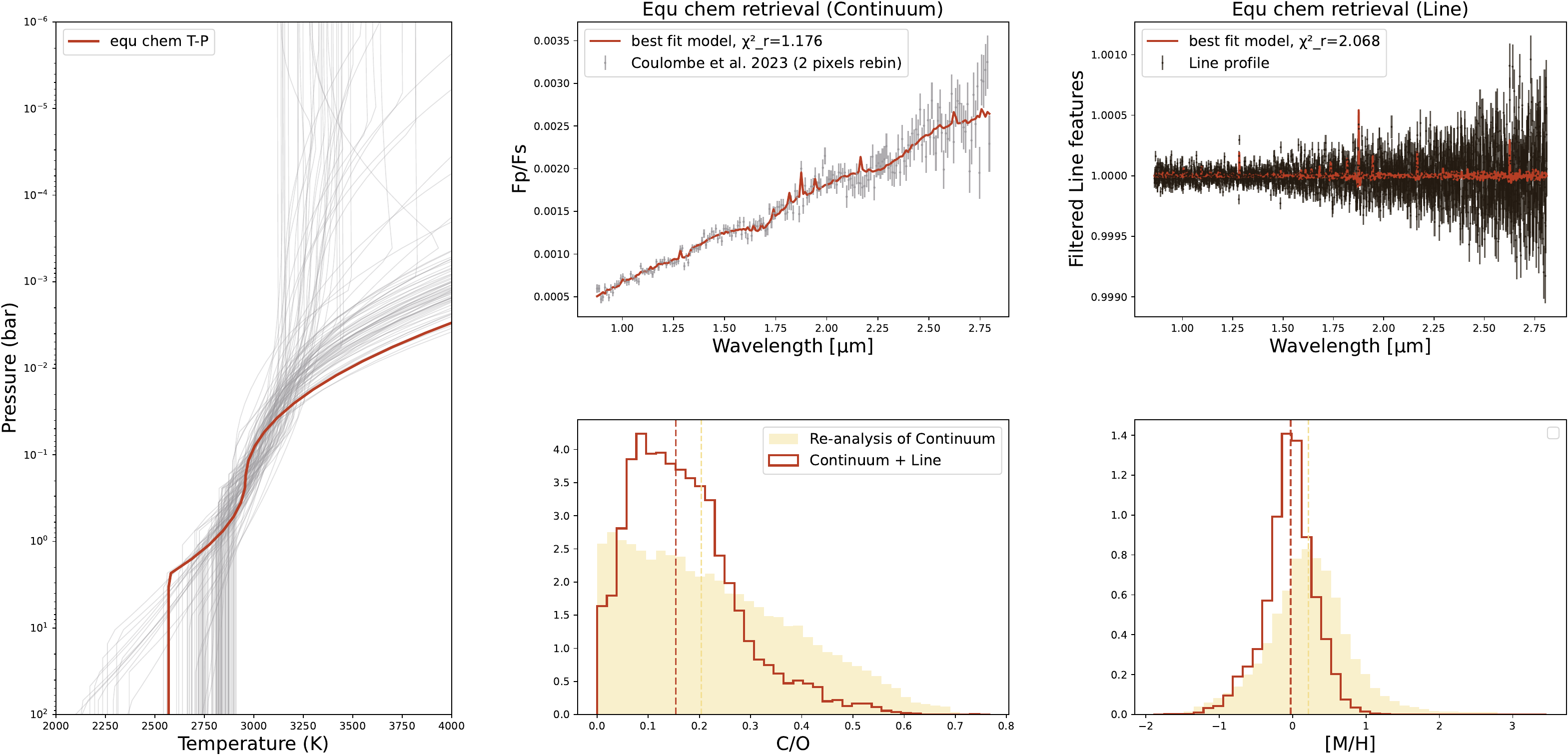}
\caption{Results of the chemical equilibrium retrieval. \textit{Left panel}: The retrieved T-P profile. \textit{Top middle and right panels}: The continuum and line profile spectra with the best-fit chemical equilibrium model, respectively. \textit{Bottom middle and right panels}: Comparison of the C/O and [M/H] posteriors from our joint retrieval and continuum-only retrieval.
\label{fig:retrieval_equ}}
\end{figure*}

\begin{figure*}[h!]
\plotone{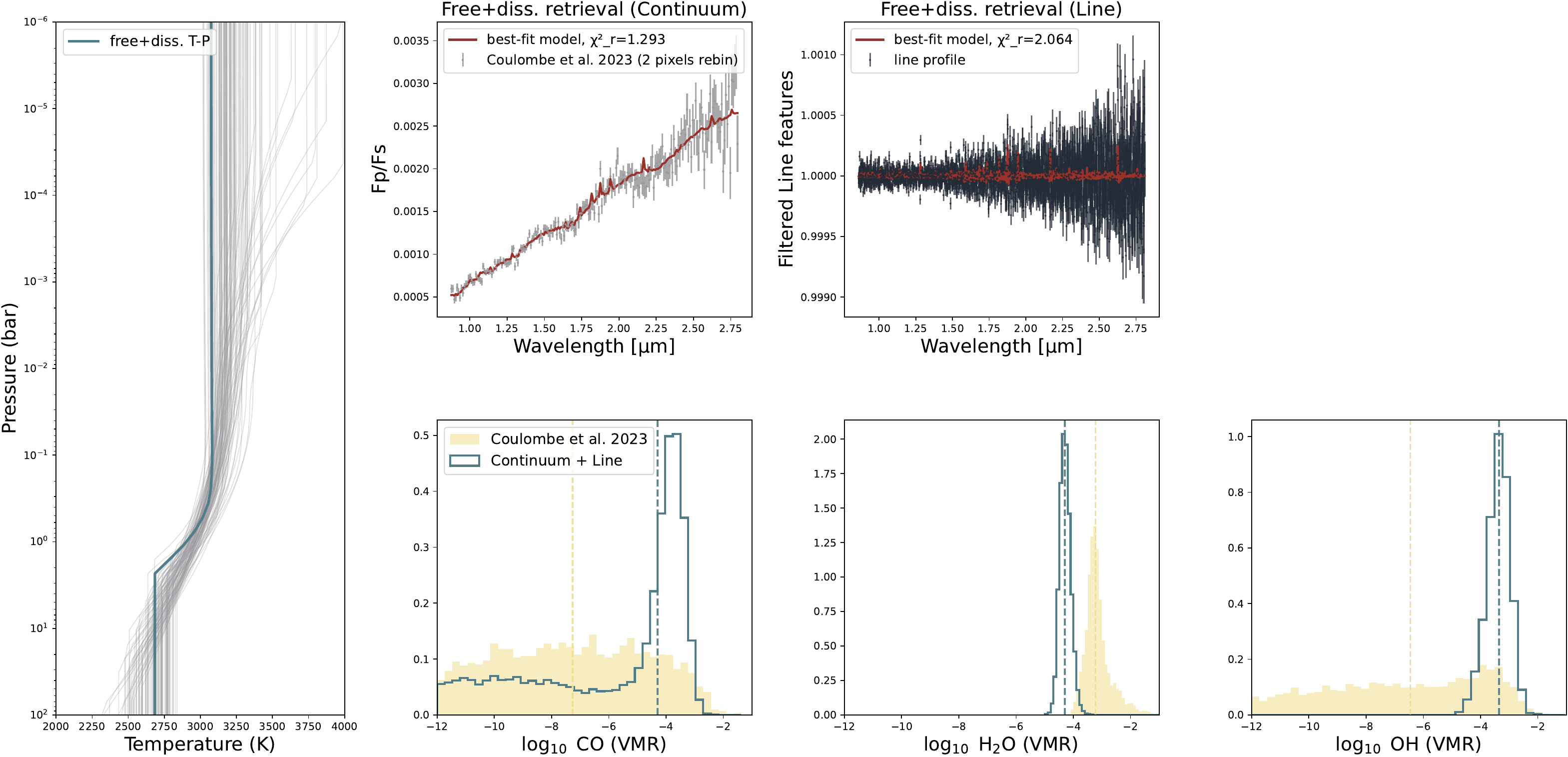}
\caption{Same as Fig. \ref{fig:retrieval_equ}, but for the free chemistry with thermal dissociation. \textit{Bottom panels}: Comparison of the retrieved CO, H$_2$O, and OH abundances from this work with the results from \citet{Coulombe2023}.
\label{fig:retrieval_free}}
\end{figure*}

The detailed retrieval settings, including parameter definitions, priors, and opacity sources, are provided in Appendix \ref{sec:retrieval results}. We performed two retrievals: one assuming chemical equilibrium and another assuming free chemistry. The retrieval results are shown in Figures \ref{fig:retrieval_equ} and \ref{fig:retrieval_free}. To facilitate a direct comparison of the results of DES+LCS and LCS only, we re-analyzed the LCS data from \citet{Coulombe2023} using our specific equilibrium chemistry parameter settings. The detailed posterior distributions of all parameters can be found in Table \ref{tab:retrieval_results}, Figures \ref{fig:corner_plot} and \ref{fig:abundance}. Our ``joint retrieval" methodology provides improved constraints on key parameters under both the chemical equilibrium and free chemistry assumptions. For the chemical equilibrium retrieval, we find C/O = $0.15^{+0.11}_{-0.09}$ and [M/H] = $-0.03^{+0.30}_{-0.32}$, which are consistent with, but provide tighter constraints than the C/O reported by \citet{Coulombe2023}. For the free+diss. retrieval, our analysis yields a posterior for the CO abundance at $\log_{10}$ VMR(CO) = $-4.30^{+0.75}_{-5.11}$. While the lower limit remains unbounded, this feature was unresolved in the \citet{Coulombe2023} analysis and is consistent with ground-based high-resolution observations \citep{Brogi2023, Yan2023}. Despite the $4.4\sigma$ in CCF detection, the unbounded lower limit of the CO posterior is likely a consequence of line-blending degeneracies at medium resolution and the unweighted nature of the joint likelihood function, which we will explore in future work. We also note a slightly lower H$_2$O abundance ($\log_{10}$ VMR(H$_2$O) = $-4.29^{+0.20}_{-0.18}$) and higher OH abundance ($\log_{10}$ VMR(OH) = $-3.36^{+0.36}_{-0.42}$), comparing to $\log_{10}$ VMR(H$_2$O) = $-3.22^{+0.46}_{-0.28}$ and non-detection of OH in \citet{Coulombe2023}. The slight difference is likely due to our medium-resolution line profiles providing information on OH opacity. The ``joint retrieval" allows the opacity within 1.3-2.7 $\mu$m to be jointly explained by both OH and H$_2$O, thereby reducing the H$_2$O abundance while simultaneously increasing the OH abundance (Fig. \ref{fig:species_contribution} \& \ref{fig:abundance}). Our constraints on H$_2$O and OH abundance are also consistent with ground-based observations of \citet{Brogi2023}.

In summary, our results demonstrate that by including the native-resolution line profile information from NIRISS, we provide the retrieval model with additional information that leads to better constraints on atmospheric parameters. This highlights the significant advantage of our methodology for analyzing JWST datasets, particularly in its ability to break model degeneracies and detect atmospheric species accurately.

\section{Discussion and Conclusions} \label{sec:dis concl}

In this work, we re-analyzed the JWST NIRISS/SOSS dataset of WASP-18b with the direct extraction method. We directly extracted a medium-resolution planetary residual spectrum at the native instrumental resolution, without performing light curve fitting. By cross-correlating the direct extraction spectrum with spectral templates, we detected emission signals of CO ($4.4 \sigma$), H$_2$O ($3.4 \sigma$), and OH ($7.8 \sigma$) in the dayside atmosphere of WASP-18b.
 
We further performed a joint retrieval analysis, combining our extracted medium-resolution spectrum with the planetary emission spectrum reported in \citet{Coulombe2023}. Our chemical equilibrium retrieval yielded C/O = $0.15^{+0.11}_{-0.09}$ and [M/H] = $-0.03^{+0.30}_{-0.32}$. The free-chemistry retrieval (including thermal dissociation) yielded $\log_{10}$ VMR(CO) = $-4.30^{+0.75}_{-5.11}$, $\log{10}$ VMR(H$_2$O) = $-4.29^{+0.20}_{-0.18}$, and $\log_{10}$ VMR(OH) = $-3.36^{+0.36}_{-0.42}$. These results all provide tighter constraints compared to those reported by \citet{Coulombe2023}.

We measured a C/O ratio of $0.15^{+0.11}_{-0.09}$ for WASP-18b, which is consistent  with the stellar C/O ratio of $0.23 \pm 0.05$ measured by \citet{Polanski2022} within 1$\sigma$, and consistent with the solar C/O ratio in 2$\sigma$. This sub-solar to solar C/O ratio disfavors formation scenario via gas accretion beyond the CO$_2$ snow line, and migrated inward after the dispersion of the protoplanetary disk, as such a scenario typically results in a carbon-rich atmosphere \citep{Oberg2011}. However, given the current precision of C/O measurements, more definitive constraints on the planetary formation pathways will require more precise measurements of the CO abundance and stellar C/O value.

Beyond the specific results for WASP-18b, the direct extraction method offers advantages over the traditional pixel-level light curve (PLLC) fitting method \citep[e.g.,][]{Feinstein2023, Schleich2025} to some extent. For example, direct extraction has broader applicability in both transit and eclipse time-series datasets, whereas PLLC fitting struggles with low S/N datasets that are difficult to model accurate transit/eclipse depths. Additionally, direct extraction naturally provides phase-resolved spectra, which allows for the precise correction of phase-dependent orbital velocity shifts. Moreover, it would be interesting to investigate whether a PLLC-based extraction and analysis performed at the pixel level on the WASP-18b/NIRISS data could demonstrate sensitivity to CO and OH similarly.

Our results thus broaden the application of cross-correlation techniques within the context of JWST datasets. As demonstrated by \citet{Esparza-Borges2023, Esparza-Borges2025}, standard CCF analysis can be used to identify molecular features in JWST NIRSpec data. We extend this technique to NIRISS/SOSS (R $\sim 500 - 1400$ ) spectral data. Notably, cross-correlation techniques excel at capturing high-frequency information in planetary spectra, whereas standard retrieval analysis remains the optimal approach for leveraging low-resolution broadband and continuum information extracted from traditional binned light curve fitting. Combining these two methodologies can further facilitate the detection of atmospheric components in a wider range of exoplanet atmospheres, helping to break the model degeneracies often unavoidable in low-resolution retrieval analysis.

Furthermore, our retrieval analysis demonstrates that combining low-resolution continuum information with medium-resolution line profile information provides more precise constraints on exoplanet atmospheres, aiding in the inference of planetary formation and evolution pathways. Such a new analytical method can be further applied to extract phase-resolved medium-resolution spectra, which will provide deeper insights into atmospheric dynamics.
%Furthermore, our retrieval analysis demonstrates that combining low-resolution continuum information with medium-resolution line profile information provides more precise constraints on exoplanet atmospheres, aiding in the inference of planetary formation and evolution pathways. 

%% Please use the acknowledgment and contribution environments. This will 
%% be anonomyized when the "anonymous" style option is used. 
\begin{acknowledgments}
We thank the anonymous referee for the thorough and useful comments and suggestions. F.Y. acknowledges the support by the National Natural Science Foundation of China (grant no. 42375118). E.P. acknowledges financial support from the Agencia Estatal de Investigaci\'on of the Ministerio de Ciencia e Innovaci\'on MCIN/AEI/10.13039/501100011033 and the ERDF “A way of making Europe” through projects PID2021-125627OB-C32 and PID2024-158486OB-C32, and from the Centre of Excellence “Severo Ochoa” award to the Instituto de Astrofisica de Canarias. W.W. acknowledges the support by the National Natural Science Foundation of China grants 62127901, National Key R\&D Program of China No. 2025YFE0102100, 2024YFA1611802 and 2025YFE0213204, the National Astronomical Observatories Chinese Academy of Sciences No. E4TQ2101, the China Manned Space Project with NO. CMS-CSST-2025-A16 and the Pre-research project on Civil Aerospace Technologies No. D010301 funded by China National Space Administration (CNSA).

This work is based on observations made with the NASA/ ESA/CSA JWST. The data were obtained from the Mikulski Archive for Space Telescopes at the Space Telescope Science Institute, which is operated by the Association of Universities for Research in Astronomy, Inc., under NASA contract NAS 5-03127 for JWST. These observations are associated with programme JWST-ERS-01366. Support for programme JWST-ERS-01366 was provided by NASA through a grant from the Space Telescope Science Institute. The data used in this work are publicly available in the Mikulski Archive for Space Telescopes (\url{https://archive.stsci.edu/}), and the specific observation can be found via \url{https://doi.org/10.17909/n6mq-s211}.

\end{acknowledgments}

% \begin{contribution}
% %%This section gives authors the space to recognize author contributions. The text inside this environment is NOT counted towards the total word quanta. At a minimum, manuscripts are expected to include this text:

% %% But authors are expected to provide more specific details, e.g. 
% %%
% %%SC was responsible for writing and submitting the manuscript.
% %%WWM came up with the initial research concept and edited the manuscript.
% %%OTS obtained the funding and edited the manuscript.
% %%EBF provided the formal analysis and validation. He also edited the manuscript.
% %%GEH Supervised the undergraduates, wrote the software and administers the project github and Zenodo repositories.
% %%
% %% Authors can use the Contributor Role Taxonomy (CRediT) at
% %% https://credit.niso.org
% %% for ideas on how write a good statement tailored to their needs.

% \end{contribution}

\facilities{JWST(NIRISS)}

\software{\texttt{astropy} \citep{Astropy_Collaboration2013, Astropy_Collaboration2018, Astropy_Collaboration2022}
            \texttt{dynesty} \citep{Speagle2020},
            \texttt{exoTEDRF} \citep{exoTEDRF, Radica2023, Feinstein2023},
            \texttt{petitRADTRANS} \citep{Molliere2019},
            \texttt{PyAstronomy} \citep{Czesla2019},
            \texttt{scipy} \citep{Virtanen2020},
          }


\appendix

\section{Wavelength solution refinement} \label{sec:wave solution}

\begin{figure*}[ht!]
\plotone{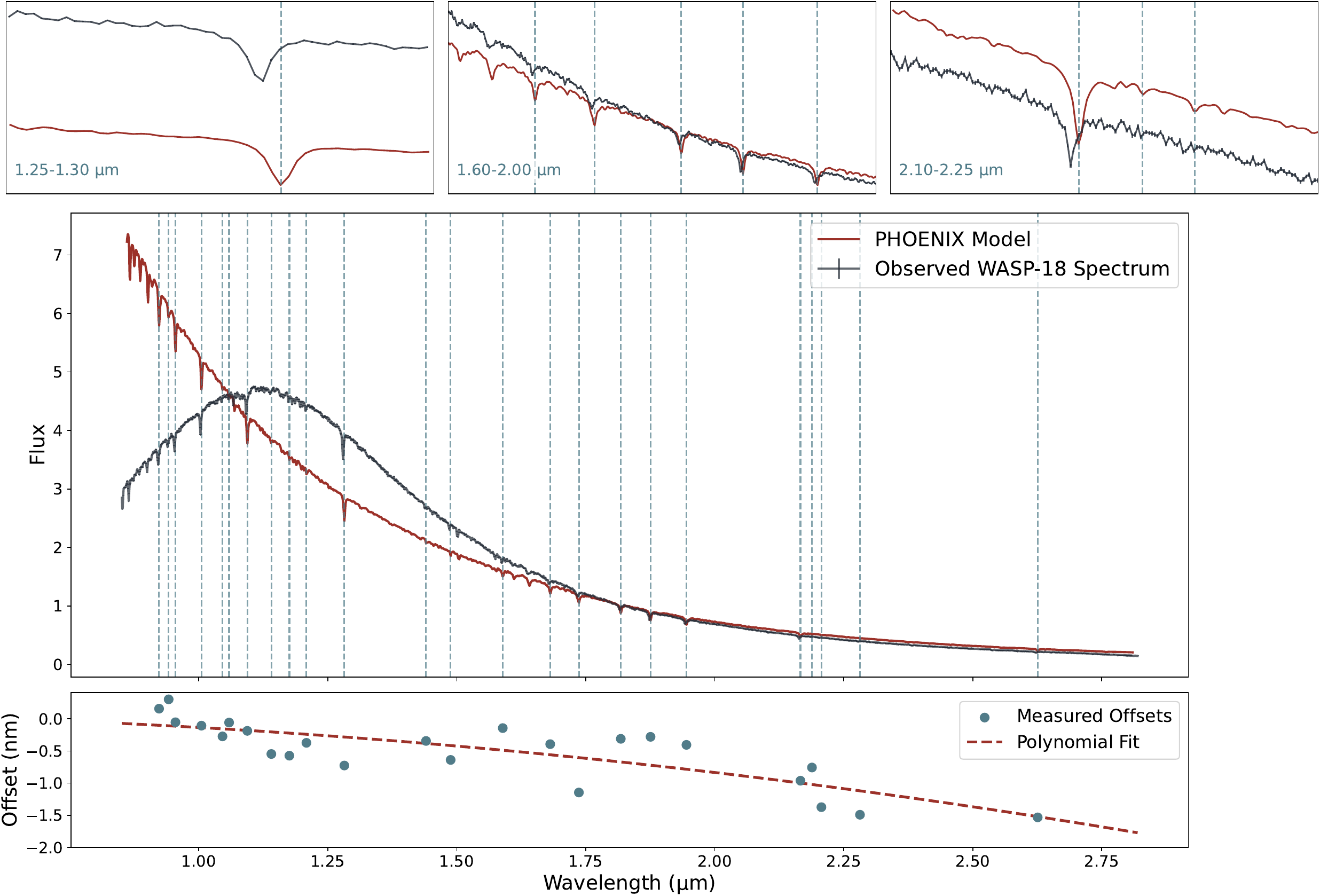}
\caption{\textit{Top panels}: Comparison of the observed WASP-18 spectrum processed by \texttt{exoTEDRF} and PHOENIX model spectrum. To compare the stellar absorption lines, both spectra have been scaled in the plot. The vertical dashed lines indicate the precise wavelengths of the stellar lines used for our wavelength optimization. The three inset panels provide zoomed-in views of the 1.20-1.30 µm, 1.60-2.00 µm, and 2.10-2.25 µm regions, respectively. A significant offset is clearly visible between the wavelength solution of the observed spectrum and the model. \textit{Bottom panels}: The offsets between the PHOENIX wavelength solution and \texttt{exoTEDRF} pipeline wavelength solution, and the polynomial fitting of these offsets.
\label{fig:steller_spec}}
\end{figure*}

We found the wavelength solution of \texttt{exoTEDRF} to be insufficiently precise for cross-corelation analysis. Upon matching strong stellar absorption lines between the observed spectra and the PHOENIX model, we noted a significant offset, varying from 0.3 to 1.5 nm (see Fig. \ref{fig:steller_spec}). We then optimized the wavelength solution by matching these stellar absorption lines.

Specifically, we first applied an identical Gaussian filter to both the observed spectra and the high-resolution PHOENIX model to isolate the absorption lines. We selected 24 prominent stellar absorption lines within the Order 1 wavelength range for this calibration. We determined the precise ``true" line cores by fitting the profiles in the model spectrum. Subsequently, we fitted the same lines in our observed spectra to determine their ``observed" line-center positions. We then computed the difference between these two sets of line cores and fitted this wavelength-dependent offset using a second-order polynomial. This polynomial, which represents the correction term, was then added to the \texttt{exoTEDRF} wavelength solution to produce our final, optimized calibration.

\section{Spectral Convolution Settings} \label{sec:convolution}

\begin{figure}[h!]
\plotone{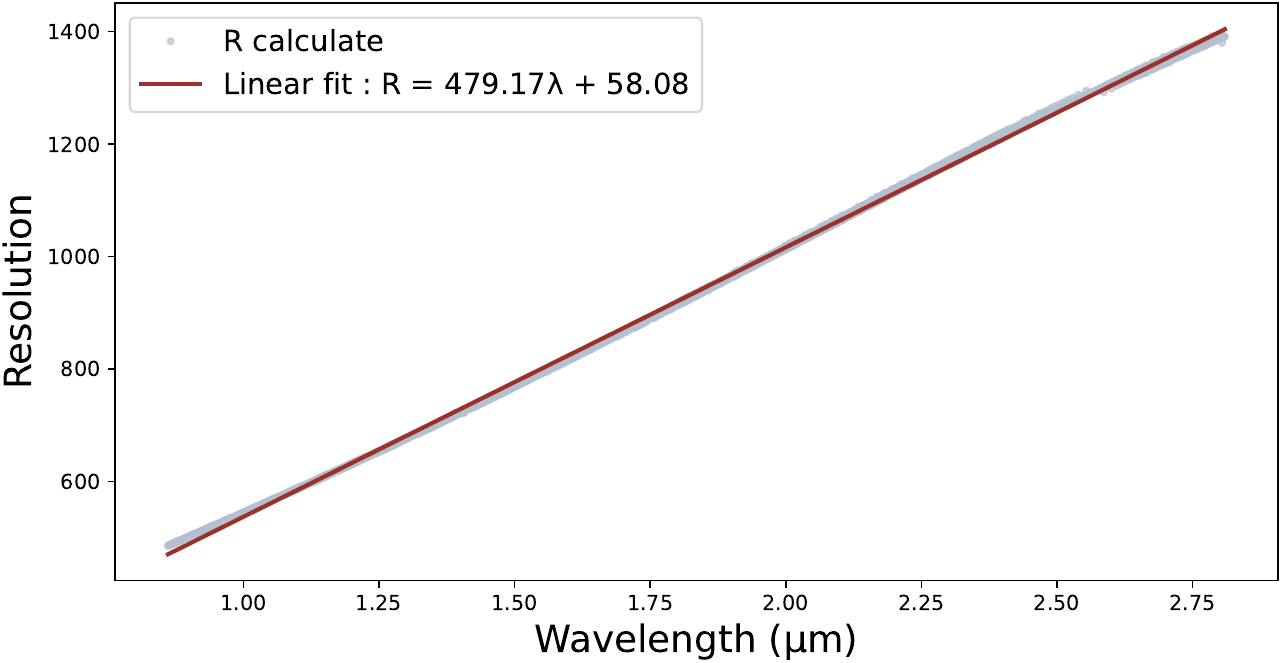}
\caption{The calculated resolution of NIRISS/SOSS (blue dots) and the linear fit resolution.
\label{fig:resolution}}
\end{figure}

The Single Object Slitless Spectroscopy (SOSS) mode of NIRISS is one of the primary instruments on JWST for exoplanetary research, designed to simultaneously cover the wavelength range from 0.6 to 2.8 $\mu$m at a resolving power of $R \sim 700$ \citep{NIRISS_intro}. However, according to Figure 7 and Equation 1 of \citet{SOSS_intro}, the spectral resolution of SOSS is not constant across wavelengths. It varies approximately linearly with wavelength, reaching a resolving power of $R \sim 1200$ at 2.3 $\mu$m in Order 1 (Fig. \ref{fig:resolution}). Consequently, convolving high-resolution template and forward model spectra to the instrumental resolution of NIRISS/SOSS cannot be performed under the assumption of a constant resolving power.

To address this, we derived a fast convolution method for variable resolution. 
Such a fast spectral convolution is a necessary step to speed up the retrieval process. We made our code \texttt{fastconv_VariR} publicly available \footnote{\url{https://github.com/qlouyang/Variable-Resolution-Convolution}}.

First, following the description in \citet{SOSS_intro} and assuming a Nyquist limit of two pixels, we linearly fitted the resolving power as a function of wavelength from our wavelength solution: $R = a\lambda + b$, where $a$ and $b$ are linear coefficients. Our objective was to find a mapping function $u(\lambda)$ such that the width of a resolution element in the transformed $u$-space is a wavelength-independent constant.

We consider a differential wavelength interval $d\lambda$, which maps to the new function space as $du = \frac{du}{d\lambda} d\lambda$. This can be approximated as $\Delta u = \frac{du}{d\lambda} \Delta\lambda$.
Given that the spectral resolution element size is $\Delta \lambda = \frac{\lambda}{R}$, and substituting the variable resolution of SOSS, we have:
\begin{equation}
    \Delta \lambda = \frac{\lambda}{a\lambda+b}
\end{equation}

We aim to make the resolution element size in the new space, $\Delta u$, a constant. For simplicity, we set $\Delta u = 1$. Substituting this into the transformation equation gives:
\begin{equation}
    1 = \frac{du}{d\lambda} \left( \frac{\lambda}{a\lambda+b} \right)
\end{equation}
which yields:
\begin{equation}
    \frac{du}{d\lambda} = \frac{a\lambda+b}{\lambda} = a + \frac{b}{\lambda}
\end{equation}

Integrating both sides with respect to $\lambda$, we obtain the required mapping function:
\begin{equation}
    u(\lambda) = \int \left( a + \frac{b}{\lambda} \right) d\lambda = a\lambda + b \ln(\lambda) + C
\end{equation}
where $C$ is a constant that can be ignored. 
Under this new space $u(\lambda)$, the spectral resolution corresponds to a constant step across the whole wavelength range. This allows us to perform the convolution using the fast Fourier transform efficiently. The convolved template and forward model spectra are then transformed back to the wavelength space for the subsequent cross-correlation and retrieval analysis.

\section{Stellar Contamination in Cross-Correlation} \label{sec:stellar_contam}

\begin{figure*}[h!]
\plotone{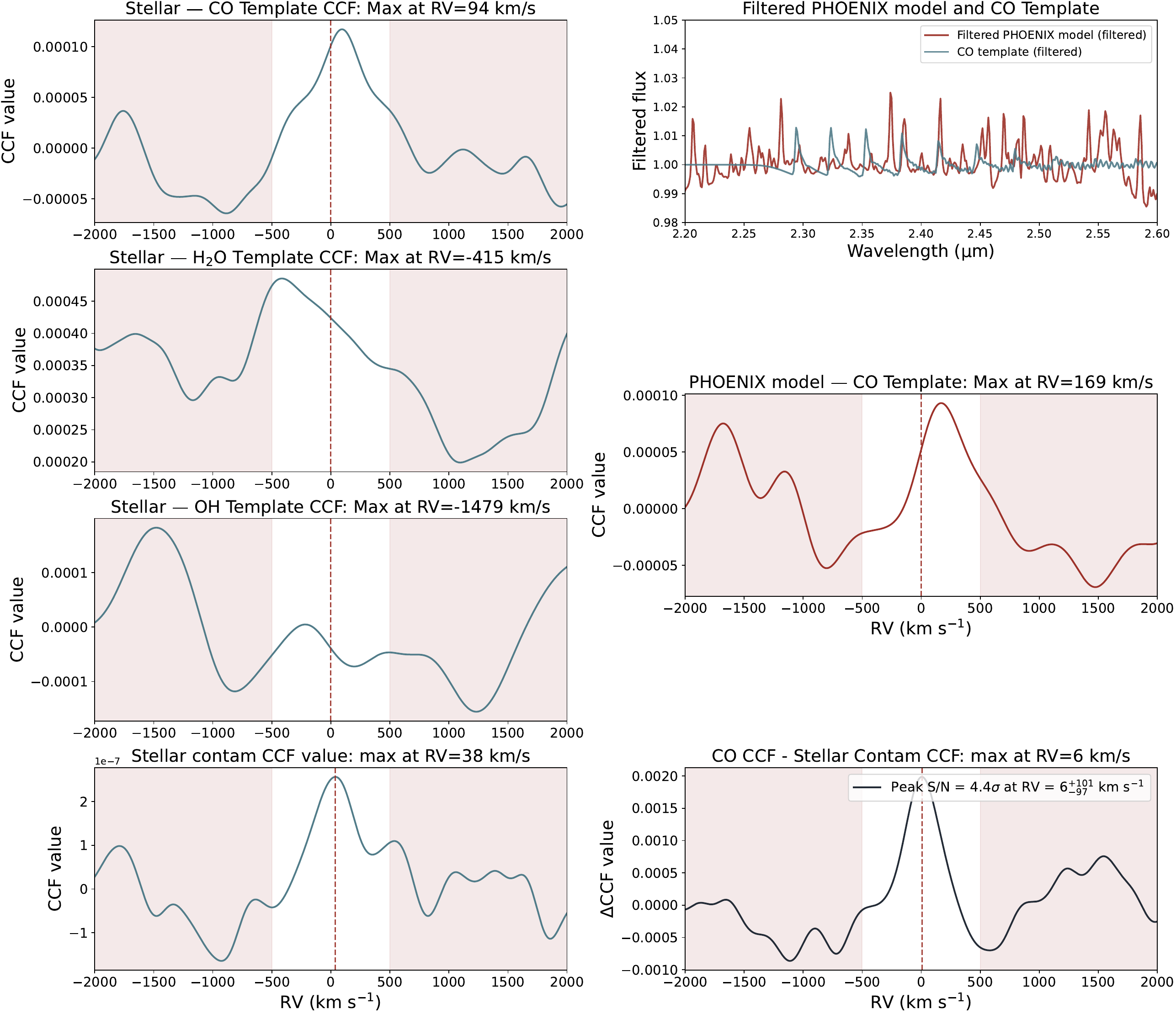}
\caption{\textit{Left three panels}: CCFs of the in-eclipse stellar spectrum of WASP-18 against the three molecular templates (CO, H$_2$O, and OH). The red vertical dashed lines denote the RV = 0 km s$^{-1}$. \textit{Top right panel}: A direct comparison between the WASP-18 PHOENIX spectrum absorption lines and the CO template emission lines in 2.2–2.6 $\mu$m. Both spectra have been scaled to facilitate visual comparison. \textit{Middle right panel}: The CCF of WASP-18 PHOENIX spectrum and the CO template, which exhibits a profile similar to the observed stellar CCF shown in the top left panel. \textit{Bottom panels}: Similar to Figure \ref{fig:temp_CCF}, the stellar contamination CCF values (left) and the planetary CO CCF subtract stellar contamination CCF value and S/N (right).
\label{fig:stellar_CCF}}
\end{figure*}

We found the peaks of H$_2$O and OH $K_p$ maps are at 0 km s$^{-1}$ (Figure. \ref{fig:Kp_maps}). To empirically rule out the stellar origin for these signals, we performed a cross-correlation analysis on the in-eclipse stellar spectrum. The stellar flux was inverted to match the emission templates. As shown in the left panels of Figure \ref{fig:stellar_CCF}, the in-eclipse stellar CCFs show no significant peaks for CO, H$_2$O, and OH at 0 km/s, confirming that these molecules are absent in the photosphere or potential starspots of WASP-18. However, we found the CO result shows a peak at 94 km s$^{-1}$. To investigate the origin of this specific feature, We compared the stellar lines and the CO template in the CO CCF wavelength range (2.2–2.6 $\mu$m), then we cross-correlated a PHOENIX stellar model spectrum with the CO template (the top right and middle panel of Figure \ref{fig:stellar_CCF}). The resulting CCF profile is similar to the stellar CCF, which determined that this offset peak is an artifact caused by a coincidental overlap between the CO template lines and certain stellar absorption lines.

Finally, we quantitatively analyzed the contribution of these specific stellar lines to the planetary CO CCF. We first constructed a ``stellar contamination spectrum": $F_{\text{contam}} = 1 + (F_{p,\text{cont}} / F_s)$, where $F_{p,\text{cont}}$ is the planetary blackbody continuum generated assuming the equilibrium temperature of WASP-18b ($T_{\text{eq}} = 2400$ K), and $F_s$ is the in-eclipse stellar spectrum. However, because our data reduction process does not involve absolute flux calibration, the uncalibrated $F_{p,\text{cont}}$ and $F_s$ cannot be directly divided. To bypass this, we constructed a mock stellar spectrum, $F_{s,\text{mock}}$, to substitute for the true $F_s$. This mock spectrum is defined as $F_{s,\text{mock}} = F_{s,\text{cont}} \times F_{s,\text{line}}$, where $F_{s,\text{cont}}$ is the stellar blackbody continuum based on the effective temperature of WASP-18 ($T_{\text{eff}} = 6400$ K), and $F_{s,\text{line}}$ is the high-pass filtered in-eclipse stellar spectrum. The final stellar contamination spectrum is thus given by:

\begin{equation}
    F_{\text{contam}} = 1 + \left( \frac{F_{p,\text{cont}}}{F_{s,\text{mock}}} \right)
\end{equation}

We then cross-correlated this $F_{\text{contam}}$ spectrum with the CO template. To isolate the pure planetary signal, the true planetary CO CCF should be:

\begin{equation}
   \text{CCF}_{\text{planet}} = \text{CCF}_{\text{original}} - \text{CCF}_{\text{contam}} 
\end{equation}

The stellar contamination CCF and the planetary CO CCF are shown in the bottom panles of Figure \ref{fig:stellar_CCF}. We found that the CCF values induced by stellar contamination are on the order of $\sim 10^{-7}$, which exerts a negligible impact on the planetary CO CCF values. Therefore, we can conclude that the detected CO signal in Figure \ref{fig:temp_CCF} \& \ref{fig:Kp_maps} is not a result of stellar contribution.

\section{Difference between DES and LCS} \label{sec:brightness}

\begin{figure*}[h!]
\plotone{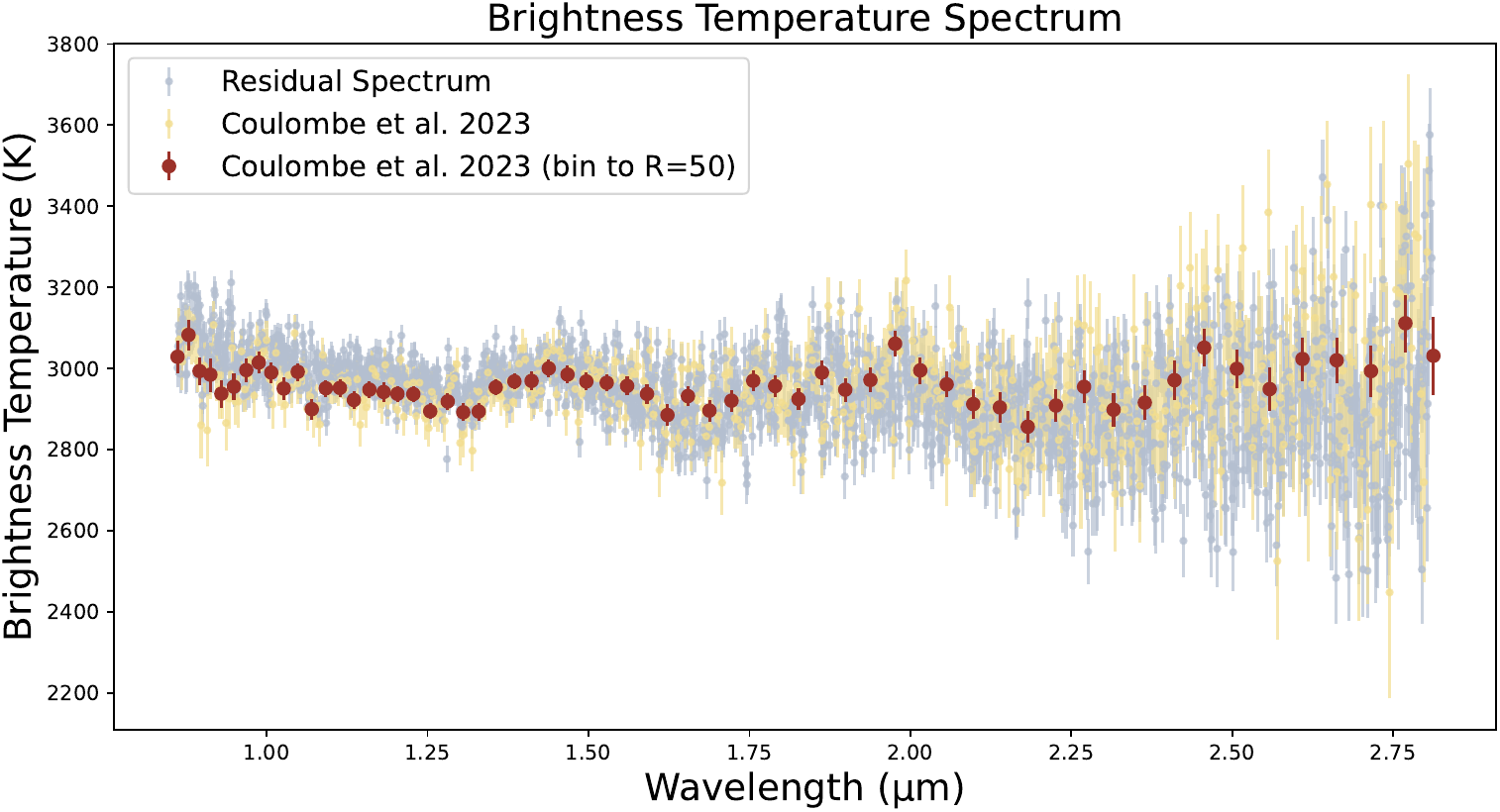}
\caption{The comparison of brightness temperatures for the direct-extracted spectrum (grey dots) and light curve fitting spectrum (yellow dots). The red dots are the same brightness temperature spectrum reported in \citet{Coulombe2023}, but bin to $R = 50$ for clarity.
\label{fig:Tb_compare}}
\end{figure*}

\begin{figure*}[h!]
\plotone{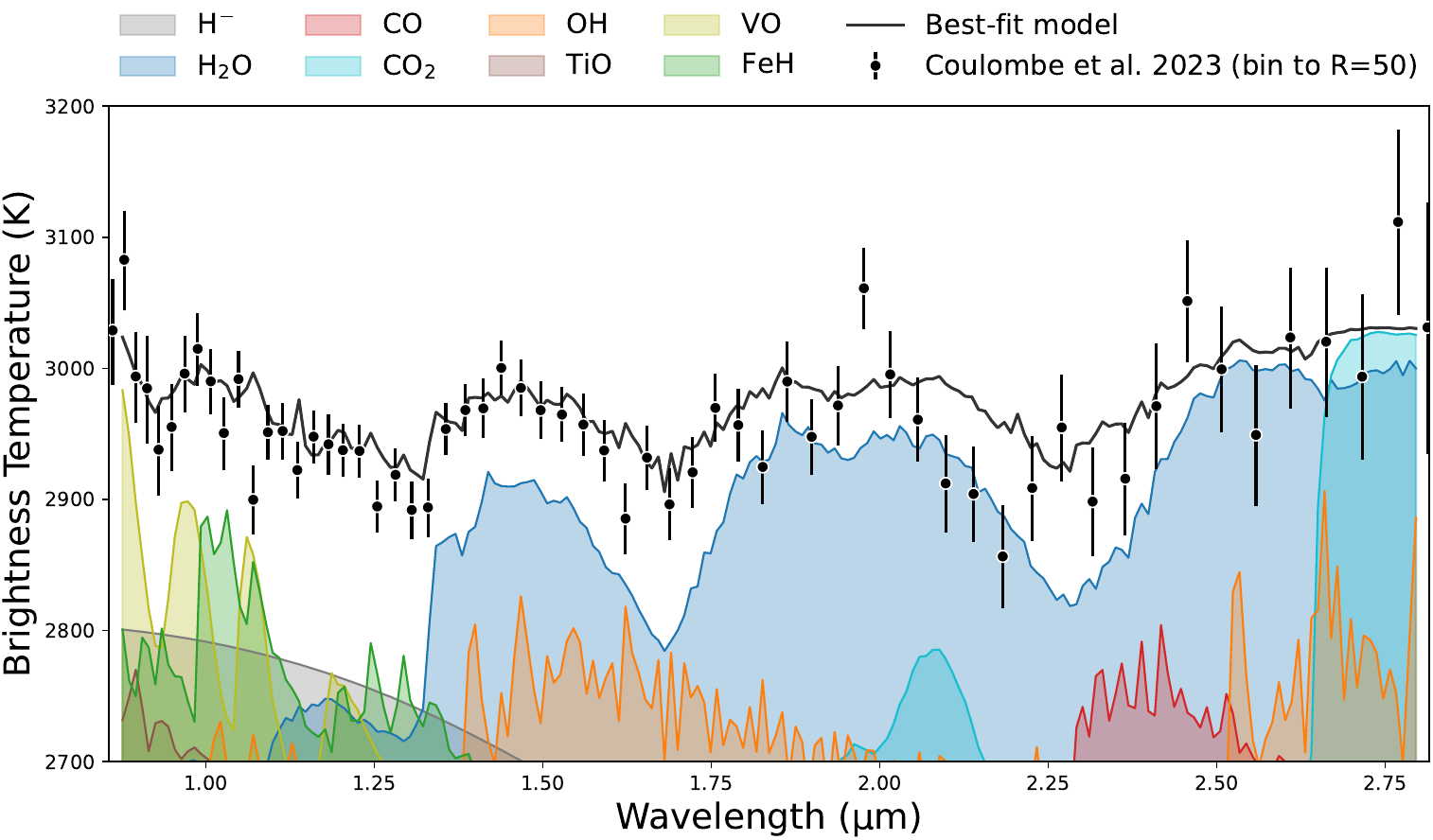}
\caption{Contribution of atmospheric species to the brightness temperature spectrum of WASP-18b, as derived from our free chemistry retrieval. The solid black line represents the best-fit model. The shaded colored regions indicate the individual contributions of different chemical species to the total opacity. The black points with error bars are the brightness temperature spectrum from \citet{Coulombe2023}.
\label{fig:species_contribution}}
\end{figure*}

We calculated the brightness temperatures for both our direct-extracted spectrum (DES) and the light curve fitting spectrum (LCS) reported by \citet{Coulombe2023}, as shown in Figure \ref{fig:Tb_compare}. We used the interpolated PHOENIX model ($T_{\mathrm{eff}}=6400$ K, $\log g=4.367$, [M/H]=0.10) to represent $F_\mathrm{s}$ to calculate $F_\mathrm{p}$ from $F_\mathrm{p} / F_\mathrm{s} + 1$. The brightness temperature of the DES is generally consistent with that of the LCS. However, subtle discrepancies are observed at the shorter and longer wavelengths. Specifically, at the blue end, the DES exhibits a slightly higher brightness temperature than the LCS, whereas at the red end, it is slightly lower. This deviation is likely attributable to an average of geometric effects and potential systematic noise. If the continuum of the residual spectrum were used directly for retrieval analysis, it could bias the abundance estimates for certain chemical species, particularly H$^-$, TiO, VO, and FeH, whose opacities dominate in the blue wavelength region (see Fig. \ref{fig:species_contribution}).

\section{Retrieval settings and detailed results} \label{sec:retrieval results}

Our retrieval framework uses \texttt{petitRADTRANS 3} to generate forward models, assuming a T-P parameterization from \citet{Madhusudhan2009}. The planetary parameters are consistent with those used for template generation in Sec. \ref{sec:CCF}, and the stellar flux used to calculate $F_\mathrm{p}/F_\mathrm{s} + 1$ is the same PHOENIX model mentioned in Sec. \ref{sec:obs reduction} and Appendix \ref{sec:brightness}. We defined an atmosphere of 50 layers, distributed uniformly in log-pressure from $10^{2}$ bar to $10^{-6}$ bar. We performed two retrieval runs: one assuming chemical equilibrium to constrain C/O and [M/H], and another assuming free chemistry to constrain the abundances of individual species. In both retrievals, we considered the same 18 chemical species mentioned in \citet{Coulombe2023}: H$_2$, H, H$^-$, e$^-$ \citep{Gray2008}, He, Na, K, Fe \citep{Kurucz2018}, H$_2$O \citep{Polyansky2018}, OH, CO, CO$_2$, CH$_4$ \citep{Yurchenko2014}, NH$_3$ \citep{Rothman2013}, HCN \citep{Harris2006}, TiO, VO, and FeH \citep{Bernath2020}. Among the 18 chemical species included in our model, five species (H$_2$, He, H, H$^-$, and e$^-$) do not contribute to line opacities. We also accounted for the Collision-Induced Absorption (CIA) of H$_2$-H$_2$ and H$_2$-He, as well as the bound-free and free-free continuum absorption of the H$^-$ ion. The abundance of e$^-$ is used for calculating the strength of H$^-$ opacity.

The chemical equilibrium retrieval (hereafter equ.) included a total of 10 free parameters. Six parameters defined the T-P profile: the temperature at the bottom of the atmosphere layer ($T_{set}$), three pressure points ($\log P_1, \log P_2, \log P_3$) to define different temperature regimes, and two free parameters $\alpha_1$ and $\alpha_2$. The retrieval also included C/O and [M/H]. In our retrieval framework, [M/H] scales the bulk metallicity and determines the oxygen abundance, and the carbon abundance is subsequently derived via the C/O ratio. We explicitly state this because [M/H] could also be linked to the carbon abundance and the final retrieved C/O value would be different. The abundances of all species were interpolated from a \texttt{FastChem 3} \citep{Kitzmann2024} chemical equilibrium grid based on the given T-P, C/O, and [M/H]. Finally, the model spectrum was computed at $R = 10,000$ and then convolved to the instrument resolution.

%We explicitly state this approach to distinguish it from alternative parameterizations where [M/H] might be anchored to the carbon abundance, as these different definitions can lead to different abundance tracks in the parameter space.

Furthermore, we fitted hotspot area fractions ($A_{HS}$), similar to \citet{Coulombe2023}. However, considering the aforementioned differences between the DES and the LCS, we assigned separate area fractions for the continuum ($A_{HS\_cont}$) and the line profiles ($A_{HS\_line}$), as $A_{HS\_line}$ also represents the phase-average information. Before the joint retrieval, we applied distinct processing to the continuum and line profile data to prevent double-fitting the same spectral information in the likelihood function. Specifically, we rebinned the \citet{Coulombe2023} continuum spectrum by 2 pixels. The original \citet{Coulombe2023} spectrum was derived by fitting light curves binned over 5 pixels. Taking into account the planetary radial velocity variation from the start to the end of the observation (from $208.53$ km s$^{-1}$ to $-165.97$ km s$^{-1}$), the planetary spectrum shifted by approximately 2 pixels across the detector (an average dispersion of $159.90$ km s$^{-1}$ per pixel for NIRISS/SOSS). Consequently, the \citet{Coulombe2023} continuum inherently contains spectral information broader than 7 pixels. By rebinning this spectrum by 2 pixels, the resulting continuum effectively isolates information on scales of $\gtrsim 14$ pixels. For the line profile spectrum, we applied a Gaussian high-pass filter with a standard deviation of $\sigma = 6$ pixels. This $\sigma$ value is sufficiently small to remove the broad continuum variations effectively, yet large enough to preserve the narrow line profile features. Notably, a Gaussian $\sigma$ of 6 pixels corresponds to a Full Width at Half Maximum of approximately 14 pixels. As a result, our high-pass filtered line profile contains spectral information exclusively on scales $\lesssim 14$ pixels. This ensures that the information from the continuum and line profile is not duplicated in the joint likelihood function.

Finally, the retrieval was implemented by evaluating the log-likelihood function using \texttt{dynesty} \citep{Speagle2020}, which conducts Nested sampling \citep{Skilling2004}. We set 1000 live points, and $\Delta \ln \mathcal{Z}_{i}$ tolerance with $0.1$.

The free-chemistry retrieval setup was similar to the chemical equilibrium run. We employed the same T-P parameterization, separate hotspot area factors ($A_{HS\_cont}$ and $A_{HS\_line}$). To account for thermal dissociation, we set the deep-atmosphere log-abundances (volume mixing ratio, VMR) of all species, excluding H$_2$, He, and H, as free parameters. The T-P dependent abundances for species affected by thermal dissociation were calculated according to Equations 1 and 2 and Table 1 of \citet{Parmentier2018}\footnote{Equation 2 in \citet{Parmentier2018} should use minus $\gamma$, not plus.}. H$_2$ and He were included as filler gases, assuming the solar abundance ratio, and the H abundance was determined entirely by the thermal dissociation of H$_2$. Subsequently, the mean molecular weight of the atmosphere was calculated based on the given abundances.

%As noted in \citet{Coulombe2023}, refractory species (e.g., TiO, VO, FeH, Na, K) are expected to have relatively low abundances. We therefore adopted their prior settings, defining $\mathcal{U}$[-12, -4] for TiO, VO, and FeH; $\mathcal{U}$[-12, -2] for Na and K; and $\mathcal{U}$[-12, -1] for all other species. In total, our free-chemistry (hereafter free+diss.) retrieval included 23 free parameters. The walkers and steps are identical to the chemical equilibrium retrieval.

\begin{figure*}[ht!]
\plotone{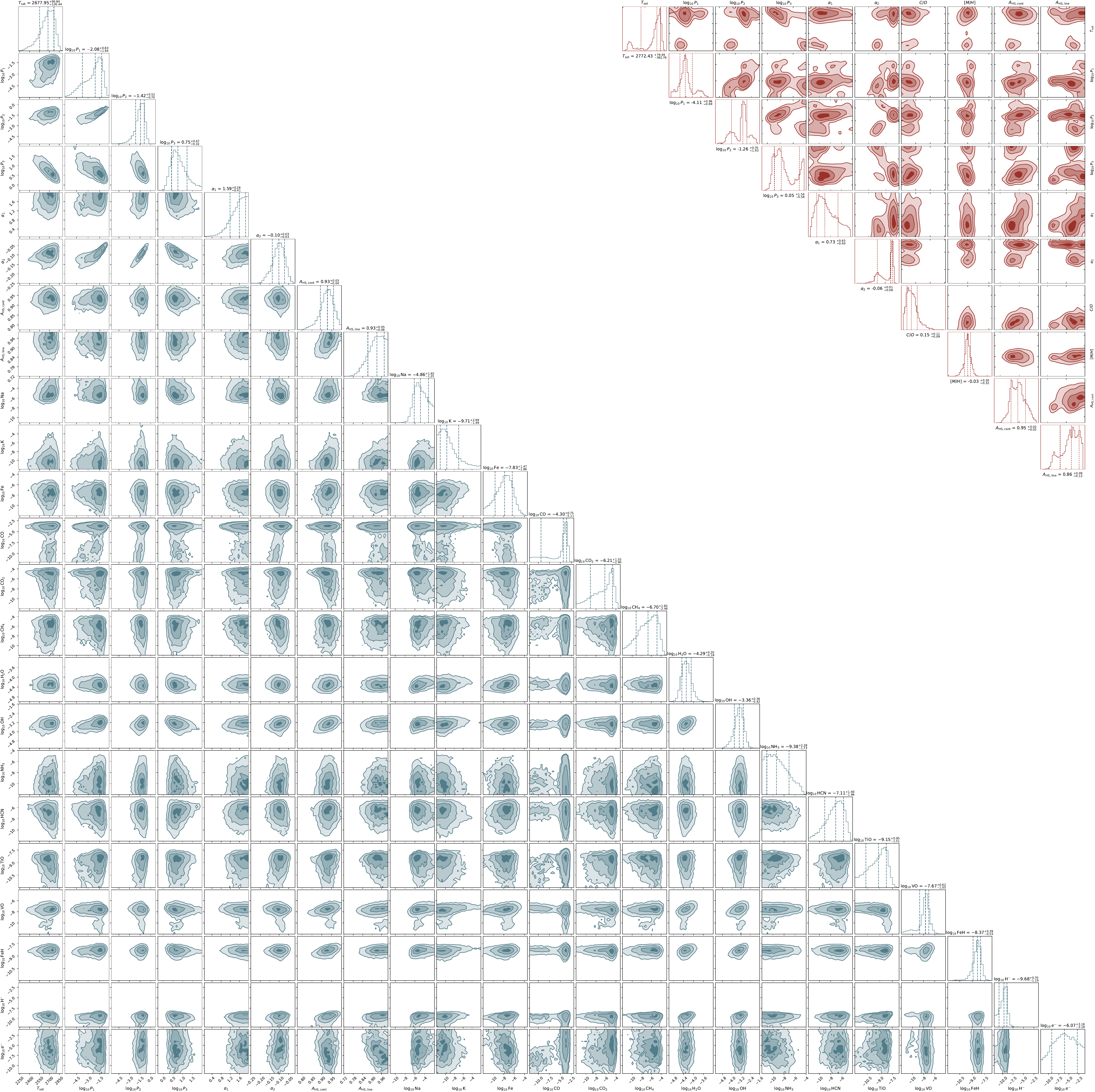}
\caption{Detailed results from the chemical equilibrium and free chemistry retrievals described in Section \ref{sec:retrieval}. \textit{Bottom Left}: Corner plot displaying the posterior distributions and correlations of the atmospheric parameters from the free chemistry retrieval. \textit{Top Right}: Corner plot displaying the posterior distributions and correlations of the atmospheric parameters from the chemical equilibrium retrieval.
\label{fig:corner_plot}}
\end{figure*}

\begin{figure*}[h!]
\plotone{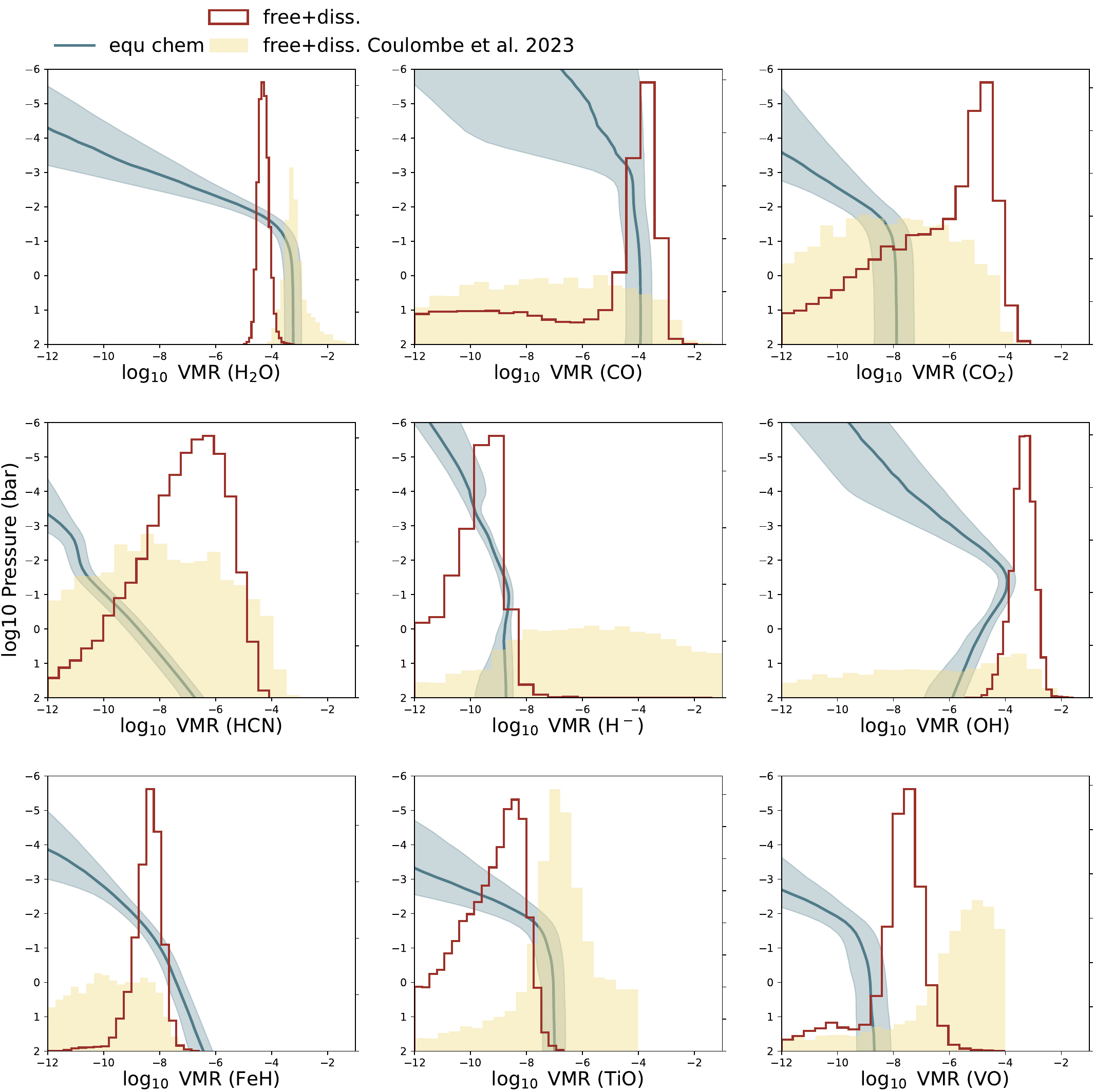}
\caption{Posterior distributions of the VMRs for key atmospheric species in WASP-18b. The blue lines and shaded regions represent the median VMR profiles and their $1\sigma$ uncertainties derived from our equilibrium chemistry retrieval. The red histograms show the posterior distributions of the deep abundances retrieved from our free chemistry with thermal dissociation (free+diss.) retrieval. For comparison, the yellow-filled histograms display the corresponding results from the free+diss. retrieval of \citet{Coulombe2023}.
\label{fig:abundance}}
\end{figure*}

The detailed posterior distribution of all parameters and the comparison can be found in Table \ref{tab:retrieval_results}, Figures \ref{fig:corner_plot} \& \ref{fig:abundance}, and the comparison of key parameters (C/O, [M/H] and the abundance of CO, H$_2$O, OH) can be found in Table \ref{tab:retrieval_results_compare}.

\begin{table}[ht!]
    \centering
    \caption{The priors and posteriors of all parameters from our retrievals.}
    \label{tab:retrieval_results}
    \renewcommand{\arraystretch}{1.5} % Increase row height for better spacing
    \begin{tabular}{lccc}
        \hline
        \hline
        Parameter & Prior & Posterior (Equilibrium Chemistry) & Posterior (Free Chemistry) \\
        \hline
        $T_{\text{set}}$ (K) & $\mathcal{U}[2,000, 4,000]$ & $2772_{-363}^{+79}$ & $2678_{-126}^{+91}$ \\
        $\log P_{1}$ (bar) & $\mathcal{U}[-6, 2]$ & $-4.11_{-0.69}^{+0.96}$ & $-2.08_{-1.64}^{+0.83}$ \\
        $\log P_{2}$ (bar) & $\mathcal{U}[-6, 2]$ & $-1.26_{-1.93}^{+0.70}$ & $-1.42_{-0.70}^{+0.52}$ \\
        $\log P_{3}$ (bar) & $\mathcal{U}[-6, 2]$ & $0.05_{-0.58}^{+1.54}$ & $0.75_{-0.32}^{+0.47}$ \\
        $\alpha_1$ & $\mathcal{U}[0.01, 2]$ & $0.73_{-0.40}^{+0.65}$ & $1.59_{-0.41}^{+0.28}$ \\
        $\alpha_2$ & $\mathcal{U}[-0.5, -0.01]$ & $-0.06_{-0.08}^{+0.01}$ & $-0.10_{-0.03}^{+0.03}$ \\
        $A_{HS\_cont}$ & $\mathcal{U}[0, 1]$ & $0.95_{-0.02}^{+0.02}$ & $0.93_{-0.04}^{+0.03}$ \\
        $A_{HS\_line}$ & $\mathcal{U}[0, 1]$ & $0.86_{-0.13}^{+0.09}$ & $0.93_{-0.06}^{+0.05}$ \\
        \hline
        C/O & $\mathcal{U}[0, 1]$ & $0.15_{-0.09}^{+0.11}$ & -- \\
        $\mathrm{[{M/H}]}$ & $\mathcal{U}[-5, 5]$ & $-0.03_{-0.32}^{+0.30}$ & -- \\
        \hline
        $\log_{10}\mathrm{VMR(Na)}$ & $\mathcal{U}[-12, -2]$ & -- & $-4.86_{-1.25}^{+1.65}$ \\
        $\log_{10}\mathrm{VMR(K)}$ & $\mathcal{U}[-12, -2]$ & -- & $-9.71_{-1.49}^{+2.69}$ \\
        $\log_{10}\mathrm{VMR(Fe)}$ & $\mathcal{U}[-12, -1]$ & -- & $-7.83_{-1.86}^{+1.47}$ \\
        $\log_{10}\mathrm{VMR(CO)}$ & $\mathcal{U}[-12, -1]$ & -- & $-4.30_{-5.11}^{+0.75}$ \\
        $\log_{10}\mathrm{VMR(CO_{2})}$ & $\mathcal{U}[-12, -1]$ & -- & $-6.21_{-2.90}^{+1.55}$ \\
        $\log_{10}\mathrm{VMR(CH_{4})}$ & $\mathcal{U}[-12, -1]$ & -- & $-6.70_{-2.46}^{+1.83}$ \\
        $\log_{10}\mathrm{VMR(H_2O)}$ & $\mathcal{U}[-12, -1]$ & -- & $-4.29_{-0.18}^{+0.20}$ \\
        $\log_{10}\mathrm{VMR(OH)}$ & $\mathcal{U}[-12, -1]$ & -- & $-3.36_{-0.42}^{+0.36}$ \\
        $\log_{10}\mathrm{VMR(NH_{3})}$ & $\mathcal{U}[-12, -1]$ & -- & $-9.38_{-1.73}^{+2.28}$ \\
        $\log_{10}\mathrm{VMR(HCN)}$ & $\mathcal{U}[-12, -1]$ & -- & $-7.11_{-1.96}^{+1.40}$ \\
        $\log_{10}\mathrm{VMR(TiO)}$ & $\mathcal{U}[-12, -4]$ & -- & $-9.15_{-1.51}^{+0.95}$ \\
        $\log_{10}\mathrm{VMR(VO)}$ & $\mathcal{U}[-12, -4]$ & -- & $-7.67_{-1.05}^{+0.62}$ \\
        $\log_{10}\mathrm{VMR(FeH)}$ & $\mathcal{U}[-12, -4]$ & -- & $-8.37_{-0.53}^{+0.39}$ \\
        $\log_{10}\mathrm{VMR(H^{-})}$ & $\mathcal{U}[-12, -1]$ & -- & $-9.68_{-1.19}^{+0.70}$ \\
        $\log_{10}\mathrm{VMR(e^{-})}$ & $\mathcal{U}[-12, -1]$ & -- & $-6.07_{-3.18}^{+3.16}$ \\
        \hline
    \end{tabular}
\end{table}

\begin{table}[ht!]
    \centering
    \caption{The comparison of key parameters posteriors from this work and previous works.}
    \label{tab:retrieval_results_compare}
    \renewcommand{\arraystretch}{1.5} % Increase row height for better spacing
    \begin{tabular}{lccc}
        \hline
        \hline
        Parameter & This work & \citet{Coulombe2023}\footnote{The C/O and $\mathrm{[{M/H}]}$ are derived from the re-analysis results of their dataset using our retrieval framework.} & \citet{Brogi2023}\footnote{The C/O and $\mathrm{[{M/H}]}$ are taken from their self-consistent 1D-RCTE grid retrieval. The abundances of three species are taken from their ``fiducial" free chemistry retrieval.} \\
        \hline
        C/O & $0.15_{-0.09}^{+0.11}$ & $0.20^{+0.20}_{-0.14}$ & $< 0.34$ \\
        $\mathrm{[{M/H}]}$ & $-0.03_{-0.32}^{+0.30}$ & $0.21^{+0.50}_{-0.52}$ & $0.48^{+0.33}_{-0.29}$ \\
        \hline
        $\log_{10}\mathrm{VMR(CO)}$ & $-4.30_{-5.11}^{+0.75}$ & -- & $> -4.12$ \\
        $\log_{10}\mathrm{VMR(H_2O)}$ & $-4.29_{-0.18}^{+0.20}$ & $-3.22^{+0.46}_{-0.28}$ & $-3.13^{+0.66}_{-0.98}$ \\
        $\log_{10}\mathrm{VMR(OH)}$ & $-3.36_{-0.42}^{+0.36}$ & -- & $> -4.74$ \\
        \hline
    \end{tabular}
\end{table}

%% For this sample we use BibTeX plus aasjournalv7.bst to generate the

\bibliography{reference}

@ARTICLE{Espinoza2025,
       author = {{Espinoza}, N{\'e}stor and {Perrin}, Marshall D.},
        title = "{Highlights from Exoplanet Observations by the James Webb Space Telescope}",
      journal = {arXiv e-prints},
         year = 2025,
        month = may,
          eid = {arXiv:2505.20520},
        pages = {arXiv:2505.20520},
          doi = {10.48550/arXiv.2505.20520},
archivePrefix = {arXiv},
       eprint = {2505.20520},
 primaryClass = {astro-ph.EP},
       adsurl = {https://ui.adsabs.harvard.edu/abs/2025arXiv250520520E}
}

@ARTICLE{Taylor2023,
       author = {{Taylor}, Jake and {Radica}, Michael and {Welbanks}, Luis and {MacDonald}, Ryan J. and {Blecic}, Jasmina and {Zamyatina}, Maria and {Roth}, Alexander and {Bean}, Jacob L. and {Parmentier}, Vivien and {Coulombe}, Louis-Philippe and {Feinstein}, Adina D. and {Espinoza}, N{\'e}stor and {Benneke}, Bj{\"o}rn and {Lafreni{\`e}re}, David and {Doyon}, Ren{\'e} and {Ahrer}, Eva-Maria},
        title = "{Awesome SOSS: atmospheric characterization of WASP-96 b using the JWST early release observations}",
      journal = {\mnras},
         year = 2023,
        month = sep,
       volume = {524},
       number = {1},
        pages = {817-834},
          doi = {10.1093/mnras/stad1547},
archivePrefix = {arXiv},
       eprint = {2305.16887},
 primaryClass = {astro-ph.EP},
       adsurl = {https://ui.adsabs.harvard.edu/abs/2023MNRAS.524..817T}
}

@ARTICLE{Ahrer2023,
       author = {{Ahrer}, Eva-Maria and {Stevenson}, Kevin B. and {Mansfield}, Megan and {Moran}, Sarah E. and {Brande}, Jonathan and {Morello}, Giuseppe and {Murray}, Catriona A. and {Nikolov}, Nikolay K. and {Petit dit de la Roche}, Dominique J.~M. and {Schlawin}, Everett and {Wheatley}, Peter J. and {Zieba}, Sebastian and {Batalha}, Natasha E. and {Damiano}, Mario and {Goyal}, Jayesh M. and {Lendl}, Monika and {Lothringer}, Joshua D. and {Mukherjee}, Sagnick and {Ohno}, Kazumasa and {Batalha}, Natalie M. and {Battley}, Matthew P. and {Bean}, Jacob L. and {Beatty}, Thomas G. and {Benneke}, Bj{\"o}rn and {Berta-Thompson}, Zachory K. and {Carter}, Aarynn L. and {Cubillos}, Patricio E. and {Daylan}, Tansu and {Espinoza}, N{\'e}stor and {Gao}, Peter and {Gibson}, Neale P. and {Gill}, Samuel and {Harrington}, Joseph and {Hu}, Renyu and {Kreidberg}, Laura and {Lewis}, Nikole K. and {Line}, Michael R. and {L{\'o}pez-Morales}, Mercedes and {Parmentier}, Vivien and {Powell}, Diana K. and {Sing}, David K. and {Tsai}, Shang-Min and {Wakeford}, Hannah R. and {Welbanks}, Luis and {Alam}, Munazza K. and {Alderson}, Lili and {Allen}, Natalie H. and {Anderson}, David R. and {Barstow}, Joanna K. and {Bayliss}, Daniel and {Bell}, Taylor J. and {Blecic}, Jasmina and {Bryant}, Edward M. and {Burleigh}, Matthew R. and {Carone}, Ludmila and {Casewell}, S.~L. and {Changeat}, Quentin and {Chubb}, Katy L. and {Crossfield}, Ian J.~M. and {Crouzet}, Nicolas and {Decin}, Leen and {D{\'e}sert}, Jean-Michel and {Feinstein}, Adina D. and {Flagg}, Laura and {Fortney}, Jonathan J. and {Gizis}, John E. and {Heng}, Kevin and {Iro}, Nicolas and {Kempton}, Eliza M.-R. and {Kendrew}, Sarah and {Kirk}, James and {Knutson}, Heather A. and {Komacek}, Thaddeus D. and {Lagage}, Pierre-Olivier and {Leconte}, J{\'e}r{\'e}my and {Lustig-Yaeger}, Jacob and {MacDonald}, Ryan J. and {Mancini}, Luigi and {May}, E.~M. and {Mayne}, N.~J. and {Miguel}, Yamila and {Mikal-Evans}, Thomas and {Molaverdikhani}, Karan and {Palle}, Enric and {Piaulet}, Caroline and {Rackham}, Benjamin V. and {Redfield}, Seth and {Rogers}, Laura K. and {Roy}, Pierre-Alexis and {Rustamkulov}, Zafar and {Shkolnik}, Evgenya L. and {Sotzen}, Kristin S. and {Taylor}, Jake and {Tremblin}, P. and {Tucker}, Gregory S. and {Turner}, Jake D. and {de Val-Borro}, Miguel and {Venot}, Olivia and {Zhang}, Xi},
        title = "{Early Release Science of the exoplanet WASP-39b with JWST NIRCam}",
      journal = {\nat},
         year = 2023,
        month = feb,
       volume = {614},
       number = {7949},
        pages = {653-658},
          doi = {10.1038/s41586-022-05590-4},
archivePrefix = {arXiv},
       eprint = {2211.10489},
 primaryClass = {astro-ph.EP},
       adsurl = {https://ui.adsabs.harvard.edu/abs/2023Natur.614..653A}
}

@ARTICLE{Alderson2023,
       author = {{Alderson}, Lili and {Wakeford}, Hannah R. and {Alam}, Munazza K. and {Batalha}, Natasha E. and {Lothringer}, Joshua D. and {Adams Redai}, Jea and {Barat}, Saugata and {Brande}, Jonathan and {Damiano}, Mario and {Daylan}, Tansu and {Espinoza}, N{\'e}stor and {Flagg}, Laura and {Goyal}, Jayesh M. and {Grant}, David and {Hu}, Renyu and {Inglis}, Julie and {Lee}, Elspeth K.~H. and {Mikal-Evans}, Thomas and {Ramos-Rosado}, Lakeisha and {Roy}, Pierre-Alexis and {Wallack}, Nicole L. and {Batalha}, Natalie M. and {Bean}, Jacob L. and {Benneke}, Bj{\"o}rn and {Berta-Thompson}, Zachory K. and {Carter}, Aarynn L. and {Changeat}, Quentin and {Col{\'o}n}, Knicole D. and {Crossfield}, Ian J.~M. and {D{\'e}sert}, Jean-Michel and {Foreman-Mackey}, Daniel and {Gibson}, Neale P. and {Kreidberg}, Laura and {Line}, Michael R. and {L{\'o}pez-Morales}, Mercedes and {Molaverdikhani}, Karan and {Moran}, Sarah E. and {Morello}, Giuseppe and {Moses}, Julianne I. and {Mukherjee}, Sagnick and {Schlawin}, Everett and {Sing}, David K. and {Stevenson}, Kevin B. and {Taylor}, Jake and {Aggarwal}, Keshav and {Ahrer}, Eva-Maria and {Allen}, Natalie H. and {Barstow}, Joanna K. and {Bell}, Taylor J. and {Blecic}, Jasmina and {Casewell}, Sarah L. and {Chubb}, Katy L. and {Crouzet}, Nicolas and {Cubillos}, Patricio E. and {Decin}, Leen and {Feinstein}, Adina D. and {Fortney}, Joanthan J. and {Harrington}, Joseph and {Heng}, Kevin and {Iro}, Nicolas and {Kempton}, Eliza M.-R. and {Kirk}, James and {Knutson}, Heather A. and {Krick}, Jessica and {Leconte}, J{\'e}r{\'e}my and {Lendl}, Monika and {MacDonald}, Ryan J. and {Mancini}, Luigi and {Mansfield}, Megan and {May}, Erin M. and {Mayne}, Nathan J. and {Miguel}, Yamila and {Nikolov}, Nikolay K. and {Ohno}, Kazumasa and {Palle}, Enric and {Parmentier}, Vivien and {Petit dit de la Roche}, Dominique J.~M. and {Piaulet}, Caroline and {Powell}, Diana and {Rackham}, Benjamin V. and {Redfield}, Seth and {Rogers}, Laura K. and {Rustamkulov}, Zafar and {Tan}, Xianyu and {Tremblin}, P. and {Tsai}, Shang-Min and {Turner}, Jake D. and {de Val-Borro}, Miguel and {Venot}, Olivia and {Welbanks}, Luis and {Wheatley}, Peter J. and {Zhang}, Xi},
        title = "{Early Release Science of the exoplanet WASP-39b with JWST NIRSpec G395H}",
      journal = {\nat},
         year = 2023,
        month = feb,
       volume = {614},
       number = {7949},
        pages = {664-669},
          doi = {10.1038/s41586-022-05591-3},
archivePrefix = {arXiv},
       eprint = {2211.10488},
 primaryClass = {astro-ph.EP},
       adsurl = {https://ui.adsabs.harvard.edu/abs/2023Natur.614..664A}
}

@ARTICLE{Rustamkulov2023,
       author = {{Rustamkulov}, Z. and {Sing}, D.~K. and {Mukherjee}, S. and {May}, E.~M. and {Kirk}, J. and {Schlawin}, E. and {Line}, M.~R. and {Piaulet}, C. and {Carter}, A.~L. and {Batalha}, N.~E. and {Goyal}, J.~M. and {L{\'o}pez-Morales}, M. and {Lothringer}, J.~D. and {MacDonald}, R.~J. and {Moran}, S.~E. and {Stevenson}, K.~B. and {Wakeford}, H.~R. and {Espinoza}, N. and {Bean}, J.~L. and {Batalha}, N.~M. and {Benneke}, B. and {Berta-Thompson}, Z.~K. and {Crossfield}, I.~J.~M. and {Gao}, P. and {Kreidberg}, L. and {Powell}, D.~K. and {Cubillos}, P.~E. and {Gibson}, N.~P. and {Leconte}, J. and {Molaverdikhani}, K. and {Nikolov}, N.~K. and {Parmentier}, V. and {Roy}, P. and {Taylor}, J. and {Turner}, J.~D. and {Wheatley}, P.~J. and {Aggarwal}, K. and {Ahrer}, E. and {Alam}, M.~K. and {Alderson}, L. and {Allen}, N.~H. and {Banerjee}, A. and {Barat}, S. and {Barrado}, D. and {Barstow}, J.~K. and {Bell}, T.~J. and {Blecic}, J. and {Brande}, J. and {Casewell}, S. and {Changeat}, Q. and {Chubb}, K.~L. and {Crouzet}, N. and {Daylan}, T. and {Decin}, L. and {D{\'e}sert}, J. and {Mikal-Evans}, T. and {Feinstein}, A.~D. and {Flagg}, L. and {Fortney}, J.~J. and {Harrington}, J. and {Heng}, K. and {Hong}, Y. and {Hu}, R. and {Iro}, N. and {Kataria}, T. and {Kempton}, E.~M.-R. and {Krick}, J. and {Lendl}, M. and {Lillo-Box}, J. and {Louca}, A. and {Lustig-Yaeger}, J. and {Mancini}, L. and {Mansfield}, M. and {Mayne}, N.~J. and {Miguel}, Y. and {Morello}, G. and {Ohno}, K. and {Palle}, E. and {Petit dit de la Roche}, D.~J.~M. and {Rackham}, B.~V. and {Radica}, M. and {Ramos-Rosado}, L. and {Redfield}, S. and {Rogers}, L.~K. and {Shkolnik}, E.~L. and {Southworth}, J. and {Teske}, J. and {Tremblin}, P. and {Tucker}, G.~S. and {Venot}, O. and {Waalkes}, W.~C. and {Welbanks}, L. and {Zhang}, X. and {Zieba}, S.},
        title = "{Early Release Science of the exoplanet WASP-39b with JWST NIRSpec PRISM}",
      journal = {\nat},
         year = 2023,
        month = feb,
       volume = {614},
       number = {7949},
        pages = {659-663},
          doi = {10.1038/s41586-022-05677-y},
archivePrefix = {arXiv},
       eprint = {2211.10487},
 primaryClass = {astro-ph.EP},
       adsurl = {https://ui.adsabs.harvard.edu/abs/2023Natur.614..659R}
}

@ARTICLE{Benneke2024,
       author = {{Benneke}, Bj{\"o}rn and {Roy}, Pierre-Alexis and {Coulombe}, Louis-Philippe and {Radica}, Michael and {Piaulet}, Caroline and {Ahrer}, Eva-Maria and {Pierrehumbert}, Raymond and {Krissansen-Totton}, Joshua and {Schlichting}, Hilke E. and {Hu}, Renyu and {Yang}, Jeehyun and {Christie}, Duncan and {Thorngren}, Daniel and {Young}, Edward D. and {Pelletier}, Stefan and {Knutson}, Heather A. and {Miguel}, Yamila and {Evans-Soma}, Thomas M. and {Dorn}, Caroline and {Gagnebin}, Anna and {Fortney}, Jonathan J. and {Komacek}, Thaddeus and {MacDonald}, Ryan and {Raul}, Eshan and {Cloutier}, Ryan and {Acuna}, Lorena and {Lafreni{\`e}re}, David and {Cadieux}, Charles and {Doyon}, Ren{\'e} and {Welbanks}, Luis and {Allart}, Romain},
        title = "{JWST Reveals CH$_4$, CO$_2$, and H$_2$O in a Metal-rich Miscible Atmosphere on a Two-Earth-Radius Exoplanet}",
      journal = {arXiv e-prints},
         year = 2024,
        month = mar,
          eid = {arXiv:2403.03325},
        pages = {arXiv:2403.03325},
          doi = {10.48550/arXiv.2403.03325},
archivePrefix = {arXiv},
       eprint = {2403.03325},
 primaryClass = {astro-ph.EP},
       adsurl = {https://ui.adsabs.harvard.edu/abs/2024arXiv240303325B}
}

@ARTICLE{Tsai2023,
       author = {{Tsai}, Shang-Min and {Lee}, Elspeth K.~H. and {Powell}, Diana and {Gao}, Peter and {Zhang}, Xi and {Moses}, Julianne and {H{\'e}brard}, Eric and {Venot}, Olivia and {Parmentier}, Vivien and {Jordan}, Sean and {Hu}, Renyu and {Alam}, Munazza K. and {Alderson}, Lili and {Batalha}, Natalie M. and {Bean}, Jacob L. and {Benneke}, Bj{\"o}rn and {Bierson}, Carver J. and {Brady}, Ryan P. and {Carone}, Ludmila and {Carter}, Aarynn L. and {Chubb}, Katy L. and {Inglis}, Julie and {Leconte}, J{\'e}r{\'e}my and {Line}, Michael and {L{\'o}pez-Morales}, Mercedes and {Miguel}, Yamila and {Molaverdikhani}, Karan and {Rustamkulov}, Zafar and {Sing}, David K. and {Stevenson}, Kevin B. and {Wakeford}, Hannah R. and {Yang}, Jeehyun and {Aggarwal}, Keshav and {Baeyens}, Robin and {Barat}, Saugata and {de Val-Borro}, Miguel and {Daylan}, Tansu and {Fortney}, Jonathan J. and {France}, Kevin and {Goyal}, Jayesh M. and {Grant}, David and {Kirk}, James and {Kreidberg}, Laura and {Louca}, Amy and {Moran}, Sarah E. and {Mukherjee}, Sagnick and {Nasedkin}, Evert and {Ohno}, Kazumasa and {Rackham}, Benjamin V. and {Redfield}, Seth and {Taylor}, Jake and {Tremblin}, Pascal and {Visscher}, Channon and {Wallack}, Nicole L. and {Welbanks}, Luis and {Youngblood}, Allison and {Ahrer}, Eva-Maria and {Batalha}, Natasha E. and {Behr}, Patrick and {Berta-Thompson}, Zachory K. and {Blecic}, Jasmina and {Casewell}, S.~L. and {Crossfield}, Ian J.~M. and {Crouzet}, Nicolas and {Cubillos}, Patricio E. and {Decin}, Leen and {D{\'e}sert}, Jean-Michel and {Feinstein}, Adina D. and {Gibson}, Neale P. and {Harrington}, Joseph and {Heng}, Kevin and {Henning}, Thomas and {Kempton}, Eliza M.-R. and {Krick}, Jessica and {Lagage}, Pierre-Olivier and {Lendl}, Monika and {Lothringer}, Joshua D. and {Mansfield}, Megan and {Mayne}, N.~J. and {Mikal-Evans}, Thomas and {Palle}, Enric and {Schlawin}, Everett and {Shorttle}, Oliver and {Wheatley}, Peter J. and {Yurchenko}, Sergei N.},
        title = "{Photochemically produced SO$_{2}$ in the atmosphere of WASP-39b}",
      journal = {\nat},
         year = 2023,
        month = may,
       volume = {617},
       number = {7961},
        pages = {483-487},
          doi = {10.1038/s41586-023-05902-2},
archivePrefix = {arXiv},
       eprint = {2211.10490},
 primaryClass = {astro-ph.EP},
       adsurl = {https://ui.adsabs.harvard.edu/abs/2023Natur.617..483T}
}

@ARTICLE{Gressier2025,
       author = {{Gressier}, Am{\'e}lie and {Batalha}, Natasha E. and {Wogan}, Nicholas and {Alderson}, Lili and {Doud}, Dominic and {Espinoza}, N{\'e}stor and {MacDonald}, Ryan J. and {Wakeford}, Hannah R. and {Valenti}, Jeff A. and {Lewis}, Nikole K. and {Seager}, Sara and {Stevenson}, Kevin B. and {Allen}, Natalie H. and {Ca{\~n}as}, Caleb I. and {Challener}, Ryan C. and {Glidden}, Ana and {Huang}, Jingcheng and {Lin}, Zifan and {Louie}, Dana R. and {Maguire}, Cathal and {Mullens}, Elijah and {Sotzen}, Kristin and {Valentine}, Daniel and {Clampin}, Mark and {Pueyo}, Laurent and {van der Marel}, Roeland P. and {Mountain}, C. Matt},
        title = "{JWST-TST DREAMS: Sulfur Dioxide in the Atmosphere of the Neptune-mass Planet HAT-P-26 b from NIRSpec G395H Transmission Spectroscopy}",
      journal = {\aj},
         year = 2025,
        month = nov,
       volume = {170},
       number = {5},
          eid = {292},
        pages = {292},
          doi = {10.3847/1538-3881/ae0929},
archivePrefix = {arXiv},
       eprint = {2509.16082},
 primaryClass = {astro-ph.EP},
       adsurl = {https://ui.adsabs.harvard.edu/abs/2025AJ....170..292G}
}

@ARTICLE{Gapp2025,
       author = {{Gapp}, Cyril and {Evans-Soma}, Thomas M. and {Barstow}, Joanna K. and {Lothringer}, Joshua D. and {Sing}, David K. and {Ruseva}, Djemma and {Ahrer}, Eva-Maria and {Goyal}, Jayesh M. and {Christie}, Duncan and {Kreidberg}, Laura and {Mayne}, Nathan J.},
        title = "{WASP-121 b's Transmission Spectrum Observed with JWST/NIRSpec G395H Reveals Thermal Dissociation and SiO in the Atmosphere}",
      journal = {\aj},
         year = 2025,
        month = jun,
       volume = {169},
       number = {6},
          eid = {341},
        pages = {341},
          doi = {10.3847/1538-3881/ad9c6e},
archivePrefix = {arXiv},
       eprint = {2506.02199},
 primaryClass = {astro-ph.EP},
       adsurl = {https://ui.adsabs.harvard.edu/abs/2025AJ....169..341G}
}

@ARTICLE{Pelletier2026,
       author = {{Pelletier}, S. and {Coulombe}, L.-P. and {Splinter}, J. and {Benneke}, B. and {MacDonald}, R.~J. and {Lafreni{\`e}re}, D. and {Cowan}, N.~B. and {Allart}, R. and {Rauscher}, E. and {Frazier}, R.~C. and {Meyer}, M.~R. and {Albert}, L. and {Dang}, L. and {Doyon}, R. and {Ehrenreich}, D. and {Flagg}, L. and {Johnstone}, D. and {Langeveld}, A.~B. and {Lim}, O. and {Piaulet-Ghorayeb}, C. and {Radica}, M. and {Rowe}, J. and {Taylor}, J. and {Turner}, J.~D.},
        title = "{Enriched volatiles and refractories but deficient titanium on the day-side atmosphere of WASP-121b revealed by JWST/NIRISS}",
      journal = {\aap},
         year = 2026,
        month = jan,
       volume = {706},
          eid = {A2},
        pages = {A2},
          doi = {10.1051/0004-6361/202556985},
archivePrefix = {arXiv},
       eprint = {2508.18341},
 primaryClass = {astro-ph.EP},
       adsurl = {https://ui.adsabs.harvard.edu/abs/2026A&A...706A...2P}
}

@ARTICLE{Grant2023,
       author = {{Grant}, David and {Lewis}, Nikole K. and {Wakeford}, Hannah R. and {Batalha}, Natasha E. and {Glidden}, Ana and {Goyal}, Jayesh and {Mullens}, Elijah and {MacDonald}, Ryan J. and {May}, Erin M. and {Seager}, Sara and {Stevenson}, Kevin B. and {Valenti}, Jeff A. and {Visscher}, Channon and {Alderson}, Lili and {Allen}, Natalie H. and {Ca{\~n}as}, Caleb I. and {Col{\'o}n}, Knicole and {Clampin}, Mark and {Espinoza}, N{\'e}stor and {Gressier}, Am{\'e}lie and {Huang}, Jingcheng and {Lin}, Zifan and {Long}, Douglas and {Louie}, Dana R. and {Pe{\~n}a-Guerrero}, Maria and {Ranjan}, Sukrit and {Sotzen}, Kristin S. and {Valentine}, Daniel and {Anderson}, Jay and {Balmer}, William O. and {Bellini}, Andrea and {Hoch}, Kielan K.~W. and {Kammerer}, Jens and {Libralato}, Mattia and {Mountain}, C. Matt and {Perrin}, Marshall D. and {Pueyo}, Laurent and {Rickman}, Emily and {Rebollido}, Isabel and {Sohn}, Sangmo Tony and {van der Marel}, Roeland P. and {Watkins}, Laura L.},
        title = "{JWST-TST DREAMS: Quartz Clouds in the Atmosphere of WASP-17b}",
      journal = {\apjl},
         year = 2023,
        month = oct,
       volume = {956},
       number = {2},
          eid = {L32},
        pages = {L32},
          doi = {10.3847/2041-8213/acfc3b10.3847/2041-8213/acfdab},
archivePrefix = {arXiv},
       eprint = {2310.08637},
 primaryClass = {astro-ph.EP},
       adsurl = {https://ui.adsabs.harvard.edu/abs/2023ApJ...956L..32G}
}

@ARTICLE{Inglis2024,
       author = {{Inglis}, Julie and {Batalha}, Natasha E. and {Lewis}, Nikole K. and {Kataria}, Tiffany and {Knutson}, Heather A. and {Kilpatrick}, Brian M. and {Gagnebin}, Anna and {Mukherjee}, Sagnick and {Pettyjohn}, Maria M. and {Crossfield}, Ian J.~M. and {Foote}, Trevor O. and {Grant}, David and {Henry}, Gregory W. and {Lally}, Maura and {McKemmish}, Laura K. and {Sing}, David K. and {Wakeford}, Hannah R. and {Zapata Trujillo}, Juan C. and {Zellem}, Robert T.},
        title = "{Quartz Clouds in the Dayside Atmosphere of the Quintessential Hot Jupiter HD 189733 b}",
      journal = {\apjl},
         year = 2024,
        month = oct,
       volume = {973},
       number = {2},
          eid = {L41},
        pages = {L41},
          doi = {10.3847/2041-8213/ad725e},
archivePrefix = {arXiv},
       eprint = {2409.11395},
 primaryClass = {astro-ph.EP},
       adsurl = {https://ui.adsabs.harvard.edu/abs/2024ApJ...973L..41I}
}

@ARTICLE{Evans-Soma2025,
       author = {{Evans-Soma}, Thomas M. and {Sing}, David K. and {Barstow}, Joanna K. and {Piette}, Anjali A.~A. and {Taylor}, Jake and {Lothringer}, Joshua D. and {Reggiani}, Henrique and {Goyal}, Jayesh M. and {Ahrer}, Eva-Maria and {Mayne}, Nathan J. and {Rustamkulov}, Zafar and {Kataria}, Tiffany and {Christie}, Duncan A. and {Gapp}, Cyril and {Dong}, Jiayin and {Foreman-Mackey}, Daniel and {Hattori}, Soichiro and {Marley}, Mark S.},
        title = "{SiO and a super-stellar C/O ratio in the atmosphere of the giant exoplanet WASP-121 b}",
      journal = {Nature Astronomy},
         year = 2025,
        month = jun,
       volume = {9},
        pages = {845-861},
          doi = {10.1038/s41550-025-02513-x},
archivePrefix = {arXiv},
       eprint = {2506.01771},
 primaryClass = {astro-ph.EP},
       adsurl = {https://ui.adsabs.harvard.edu/abs/2025NatAs...9..845E}
}

@ARTICLE{Krishnamurthy2025,
       author = {{Krishnamurthy}, Vigneshwaran and {Carteret}, Yann and {Piaulet-Ghorayeb}, Caroline and {Splinter}, Jared and {Doshi}, Dhvani and {Radica}, Michael and {Coulombe}, Louis-Philippe and {Allart}, Romain and {Bourrier}, Vincent and {Cowan}, Nicolas B. and {Lafreni{\`e}re}, David and {Albert}, Lo{\"\i}c and {Dang}, Lisa and {Jayawardhana}, Ray and {Johnstone}, Doug and {Kaltenegger}, Lisa and {Langeveld}, Adam B. and {Pelletier}, Stefan and {Rowe}, Jason F. and {Roy}, Pierre-Alexis and {Taylor}, Jake and {Turner}, Jake D.},
        title = "{Continuous helium absorption from the leading and trailing tails of WASP-107b}",
      journal = {arXiv e-prints},
         year = 2025,
        month = may,
          eid = {arXiv:2505.20588},
        pages = {arXiv:2505.20588},
          doi = {10.48550/arXiv.2505.20588},
archivePrefix = {arXiv},
       eprint = {2505.20588},
 primaryClass = {astro-ph.EP},
       adsurl = {https://ui.adsabs.harvard.edu/abs/2025arXiv250520588K}
}

@ARTICLE{Welbanks2025,
       author = {{Welbanks}, Luis and {Nixon}, Matthew C. and {McGill}, Peter and {Tilke}, Lana J. and {Wiser}, Lindsey S. and {Rotman}, Yoav and {Mukherjee}, Sagnick and {Feinstein}, Adina and {Line}, Michael R. and {Benneke}, Bj{\"o}rn and {Seager}, Sara and {Beatty}, Thomas G. and {Seligman}, Darryl Z. and {Parmentier}, Vivien and {Sing}, David},
        title = "{The Challenges Involved in the Detection of Gases in Exoplanet Atmospheres}",
      journal = {arXiv e-prints},
         year = 2025,
        month = apr,
          eid = {arXiv:2504.21788},
        pages = {arXiv:2504.21788},
          doi = {10.48550/arXiv.2504.21788},
archivePrefix = {arXiv},
       eprint = {2504.21788},
 primaryClass = {astro-ph.EP},
       adsurl = {https://ui.adsabs.harvard.edu/abs/2025arXiv250421788W}
}

@ARTICLE{Kipping2025,
       author = {{Kipping}, David and {Benneke}, Bj{\"o}rn},
        title = "{Exoplaneteers Keep Overestimating Sigma Significances}",
      journal = {arXiv e-prints},
         year = 2025,
        month = jun,
          eid = {arXiv:2506.05392},
        pages = {arXiv:2506.05392},
          doi = {10.48550/arXiv.2506.05392},
archivePrefix = {arXiv},
       eprint = {2506.05392},
 primaryClass = {astro-ph.IM},
       adsurl = {https://ui.adsabs.harvard.edu/abs/2025arXiv250605392K}
}

@ARTICLE{Thorngren2025,
       author = {{Thorngren}, Daniel P. and {Sing}, David K. and {Mukherjee}, Sagnick},
        title = "{Bayesian Model Comparison and Significance: Widespread Errors and how to Correct Them}",
      journal = {arXiv e-prints},
         year = 2025,
        month = sep,
          eid = {arXiv:2510.00169},
        pages = {arXiv:2510.00169},
          doi = {10.48550/arXiv.2510.00169},
archivePrefix = {arXiv},
       eprint = {2510.00169},
 primaryClass = {astro-ph.EP},
       adsurl = {https://ui.adsabs.harvard.edu/abs/2025arXiv251000169T}
}

@ARTICLE{Esparza-Borges2023,
       author = {{Esparza-Borges}, Emma and {L{\'o}pez-Morales}, Mercedes and {Adams Redai}, J{\'e}a I. and {Pall{\'e}}, Enric and {Kirk}, James and {Casasayas-Barris}, N{\'u}ria and {Batalha}, Natasha E. and {Rackham}, Benjamin V. and {Bean}, Jacob L. and {Casewell}, S.~L. and {Decin}, Leen and {Dos Santos}, Leonardo A. and {Garc{\'\i}a Mu{\~n}oz}, Antonio and {Harrington}, Joseph and {Heng}, Kevin and {Hu}, Renyu and {Mancini}, Luigi and {Molaverdikhani}, Karan and {Morello}, Giuseppe and {Nikolov}, Nikolay K. and {Nixon}, Matthew C. and {Redfield}, Seth and {Stevenson}, Kevin B. and {Wakeford}, Hannah R. and {Alam}, Munazza K. and {Benneke}, Bj{\"o}rn and {Blecic}, Jasmina and {Crouzet}, Nicolas and {Daylan}, Tansu and {Inglis}, Julie and {Kreidberg}, Laura and {Petit dit de la Roche}, Dominique J.~M. and {Turner}, Jake D.},
        title = "{Detection of Carbon Monoxide in the Atmosphere of WASP-39b Applying Standard Cross-correlation Techniques to JWST NIRSpec G395H Data}",
      journal = {\apjl},
         year = 2023,
        month = sep,
       volume = {955},
       number = {1},
          eid = {L19},
        pages = {L19},
          doi = {10.3847/2041-8213/acf27b},
archivePrefix = {arXiv},
       eprint = {2309.00036},
 primaryClass = {astro-ph.EP},
       adsurl = {https://ui.adsabs.harvard.edu/abs/2023ApJ...955L..19E}
}

@ARTICLE{Esparza-Borges2025,
       author = {{Esparza-Borges}, Emma and {L{\'o}pez-Morales}, Mercedes and {Pall{\'e}}, Enric and {Makhnev}, Vladimir and {Gordon}, Iouli and {Hargreaves}, Robert and {Kirk}, James and {C{\'a}ceres}, Claudio and {Crossfield}, Ian J.~M. and {Crouzet}, Nicolas and {Decin}, Leen and {D{\'e}sert}, Jean-Michel and {Flagg}, Laura and {Mu{\~n}oz}, Antonio Garc{\'\i}a and {Harrington}, Joseph and {Molaverdikhani}, Karan and {Morello}, Giuseppe and {Nikolov}, Nikolay and {Solmaz}, Arif and {Rackham}, Benjamin V. and {Redfield}, Seth},
        title = "{Testing the performance of cross-correlation techniques to search for molecular features in JWST NIRSpec G395H observations of transiting exoplanets}",
      journal = {\mnras},
         year = 2025,
        month = nov,
       volume = {543},
       number = {4},
        pages = {3456-3473},
          doi = {10.1093/mnras/staf1659},
archivePrefix = {arXiv},
       eprint = {2509.25319},
 primaryClass = {astro-ph.EP},
       adsurl = {https://ui.adsabs.harvard.edu/abs/2025MNRAS.543.3456E}
}

@ARTICLE{Zhang2025,
       author = {{Zhang}, Michael and {Beleznay}, Maya and {Brandt}, Timothy D. and {Romani}, Roger W. and {Gao}, Peter and {Beltz}, Hayley and {Bailes}, Matthew and {Nixon}, Matthew C. and {Bean}, Jacob L. and {Komacek}, Thaddeus D. and {Coy}, Brandon P. and {Fu}, Guangwei and {Luque}, Rafael and {Reardon}, Daniel J. and {Carli}, Emma and {Shannon}, Ryan M. and {Fortney}, Jonathan J. and {Piette}, Anjali A.~A. and {Miller}, M. Coleman and {Desert}, Jean-Michel},
        title = "{A Carbon-rich Atmosphere on a Windy Pulsar Planet}",
      journal = {\apjl},
         year = 2025,
        month = dec,
       volume = {995},
       number = {2},
          eid = {L64},
        pages = {L64},
          doi = {10.3847/2041-8213/ae157c},
archivePrefix = {arXiv},
       eprint = {2509.04558},
 primaryClass = {astro-ph.EP},
       adsurl = {https://ui.adsabs.harvard.edu/abs/2025ApJ...995L..64Z}
}

@ARTICLE{Schleich2025,
       author = {{Schleich}, S. and {Boro Saikia}, S. and {Change{\^a}t}, Q. and {G{\"u}del}, M. and {Voigt}, A. and {Waldmann}, I.},
        title = "{Knobs and dials of retrieving JWST transmission spectra: II. Impacts of pipeline-level differences on retrieval posteriors}",
      journal = {\aap},
         year = 2025,
        month = dec,
       volume = {704},
          eid = {A223},
        pages = {A223},
          doi = {10.1051/0004-6361/202556553},
archivePrefix = {arXiv},
       eprint = {2511.05652},
 primaryClass = {astro-ph.EP},
       adsurl = {https://ui.adsabs.harvard.edu/abs/2025A&A...704A.223S}
}

@ARTICLE{Lustig2025,
       author = {{Lustig-Yaeger}, Jacob and {Sotzen}, Kristin S. and {Stevenson}, Kevin B. and {Tsai}, Shang-Min and {Challener}, Ryan C. and {Goyal}, Jayesh and {Lewis}, Nikole K. and {Louie}, Dana R. and {Mayorga}, L.~C. and {Valentine}, Daniel and {Wakeford}, Hannah R. and {Alderson}, Lili and {Allen}, Natalie H. and {Fauchez}, Thomas J. and {Glidden}, Ana and {Gressier}, Am{\'e}lie and {H{\"o}rst}, Sarah M. and {Huang}, Jingcheng and {Lin}, Zifan and {Mandell}, Avi M. and {Mullens}, Elijah and {Peacock}, Sarah and {Schwieterman}, Edward W. and {Valenti}, Jeff A. and {Mountain}, C. Matt and {Perrin}, Marshall and {van der Marel}, Roeland P.},
        title = "{JWST-TST DREAMS: The Nightside Emission and Chemistry of WASP-17b}",
      journal = {\apjl},
         year = 2025,
        month = nov,
       volume = {994},
       number = {1},
          eid = {L4},
        pages = {L4},
          doi = {10.3847/2041-8213/ae17ae},
archivePrefix = {arXiv},
       eprint = {2510.06169},
 primaryClass = {astro-ph.EP},
       adsurl = {https://ui.adsabs.harvard.edu/abs/2025ApJ...994L...4L}
}

@ARTICLE{Snellen2010,
       author = {{Snellen}, Ignas A.~G. and {de Kok}, Remco J. and {de Mooij}, Ernst J.~W. and {Albrecht}, Simon},
        title = "{The orbital motion, absolute mass and high-altitude winds of exoplanet HD209458b}",
      journal = {\nat},
         year = 2010,
        month = jun,
       volume = {465},
       number = {7301},
        pages = {1049-1051},
          doi = {10.1038/nature09111},
archivePrefix = {arXiv},
       eprint = {1006.4364},
 primaryClass = {astro-ph.EP},
       adsurl = {https://ui.adsabs.harvard.edu/abs/2010Natur.465.1049S}
}

@ARTICLE{NIRISS_intro,
       author = {{Doyon}, Ren{\'e} and {Willott}, Chris J. and {Hutchings}, John B. and {Sivaramakrishnan}, Anand and {Albert}, Lo{\"\i}c and {Lafreni{\`e}re}, David and {Rowlands}, Neil and {Bego{\~n}a Vila}, M. and {Martel}, Andr{\'e} R. and {LaMassa}, Stephanie and {Aldridge}, David and {Artigau}, {\'E}tienne and {Cameron}, Peter and {Chayer}, Pierre and {Cook}, Neil J. and {Cooper}, Rachel A. and {Darveau-Bernier}, Antoine and {Dupuis}, Jean and {Earnshaw}, Colin and {Espinoza}, N{\'e}stor and {Filippazzo}, Joseph C. and {Fullerton}, Alexander W. and {Gaudreau}, Daniel and {Gawlik}, Roman and {Goudfrooij}, Paul and {Haley}, Craig and {Kammerer}, Jens and {Kendall}, David and {Lambros}, Scott D. and {Ignat}, Luminita Ilinca and {Maszkiewicz}, Michael and {McColgan}, Ashley and {Morishita}, Takahiro and {Ouellette}, Nathalie N.-Q. and {Pacifici}, Camilla and {Philippi}, Natasha and {Radica}, Michael and {Ravindranath}, Swara and {Rowe}, Jason and {Roy}, Arpita and {Roy}, Niladri and {Saad}, Karl and {Sohn}, Sangmo Tony and {Talens}, Geert Jan and {Touahri}, Driss and {Thatte}, Deepashri and {Taylor}, Joanna M. and {Vandal}, Thomas and {Volk}, Kevin and {Wander}, Michel and {Warner}, Gerald and {Zheng}, Sheng-Hai and {Zhou}, Julia and {Abraham}, Roberto and {Beaulieu}, Mathilde and {Benneke}, Bj{\"o}rn and {Ferrarese}, Laura and {Jayawardhana}, Ray and {Johnstone}, Doug and {Kaltenegger}, Lisa and {Meyer}, Michael R. and {Pipher}, Judy L. and {Rameau}, Julien and {Rieke}, Marcia and {Salhi}, Salma and {Sawicki}, Marcin},
        title = "{The Near Infrared Imager and Slitless Spectrograph for the James Webb Space Telescope. I. Instrument Overview and In-flight Performance}",
      journal = {\pasp},
         year = 2023,
        month = sep,
       volume = {135},
       number = {1051},
          eid = {098001},
        pages = {098001},
          doi = {10.1088/1538-3873/acd41b},
archivePrefix = {arXiv},
       eprint = {2306.03277},
 primaryClass = {astro-ph.IM},
       adsurl = {https://ui.adsabs.harvard.edu/abs/2023PASP..135i8001D}
}

@ARTICLE{SOSS_intro,
       author = {{Albert}, Lo{\"\i}c and {Lafreni{\`e}re}, David and {Ren{\'e}},, Doyon and {Artigau}, {\'E}tienne and {Volk}, Kevin and {Goudfrooij}, Paul and {Martel}, Andr{\'e} R. and {Radica}, Michael and {Rowe}, Jason and {Espinoza}, N{\'e}stor and {Roy}, Arpita and {Filippazzo}, Joseph C. and {Darveau-Bernier}, Antoine and {Talens}, Geert Jan and {Sivaramakrishnan}, Anand and {Willott}, Chris J. and {Fullerton}, Alexander W. and {LaMassa}, Stephanie and {Hutchings}, John B. and {Rowlands}, Neil and {Vila}, M. Bego{\~n}a and {Zhou}, Julia and {Aldridge}, David and {Maszkiewicz}, Michael and {Beaulieu}, Mathilde and {Cook}, Neil J. and {Piaulet}, Caroline and {Roy}, Pierre-Alexis and {Lamontagne}, Pierrot and {Morel}, Kim and {Frost}, William and {Salhi}, Salma and {Coulombe}, Louis-Philippe and {Benneke}, Bj{\"o}rn and {MacDonald}, Ryan J. and {Johnstone}, Doug and {Turner}, Jake D. and {Fournier-Tondreau}, Marylou and {Allart}, Romain and {Kaltenegger}, Lisa},
        title = "{The Near Infrared Imager and Slitless Spectrograph for the James Webb Space Telescope. III. Single Object Slitless Spectroscopy}",
      journal = {\pasp},
         year = 2023,
        month = jul,
       volume = {135},
       number = {1049},
          eid = {075001},
        pages = {075001},
          doi = {10.1088/1538-3873/acd7a3},
archivePrefix = {arXiv},
       eprint = {2306.04572},
 primaryClass = {astro-ph.IM},
       adsurl = {https://ui.adsabs.harvard.edu/abs/2023PASP..135g5001A}
}

@ARTICLE{Hellier2009,
       author = {{Hellier}, Coel and {Anderson}, D.~R. and {Collier Cameron}, A. and {Gillon}, M. and {Hebb}, L. and {Maxted}, P.~F.~L. and {Queloz}, D. and {Smalley}, B. and {Triaud}, A.~H.~M.~J. and {West}, R.~G. and {Wilson}, D.~M. and {Bentley}, S.~J. and {Enoch}, B. and {Horne}, K. and {Irwin}, J. and {Lister}, T.~A. and {Mayor}, M. and {Parley}, N. and {Pepe}, F. and {Pollacco}, D.~L. and {Segransan}, D. and {Udry}, S. and {Wheatley}, P.~J.},
        title = "{An orbital period of 0.94days for the hot-Jupiter planet WASP-18b}",
      journal = {\nat},
         year = 2009,
        month = aug,
       volume = {460},
       number = {7259},
        pages = {1098-1100},
          doi = {10.1038/nature08245},
       adsurl = {https://ui.adsabs.harvard.edu/abs/2009Natur.460.1098H}
}

@ARTICLE{Coulombe2023,
       author = {{Coulombe}, Louis-Philippe and {Benneke}, Bj{\"o}rn and {Challener}, Ryan and {Piette}, Anjali A.~A. and {Wiser}, Lindsey S. and {Mansfield}, Megan and {MacDonald}, Ryan J. and {Beltz}, Hayley and {Feinstein}, Adina D. and {Radica}, Michael and {Savel}, Arjun B. and {Dos Santos}, Leonardo A. and {Bean}, Jacob L. and {Parmentier}, Vivien and {Wong}, Ian and {Rauscher}, Emily and {Komacek}, Thaddeus D. and {Kempton}, Eliza M. -R. and {Tan}, Xianyu and {Hammond}, Mark and {Lewis}, Neil T. and {Line}, Michael R. and {Lee}, Elspeth K.~H. and {Shivkumar}, Hinna and {Crossfield}, Ian J.~M. and {Nixon}, Matthew C. and {Rackham}, Benjamin V. and {Wakeford}, Hannah R. and {Welbanks}, Luis and {Zhang}, Xi and {Batalha}, Natalie M. and {Berta-Thompson}, Zachory K. and {Changeat}, Quentin and {D{\'e}sert}, Jean-Michel and {Espinoza}, N{\'e}stor and {Goyal}, Jayesh M. and {Harrington}, Joseph and {Knutson}, Heather A. and {Kreidberg}, Laura and {L{\'o}pez-Morales}, Mercedes and {Shporer}, Avi and {Sing}, David K. and {Stevenson}, Kevin B. and {Aggarwal}, Keshav and {Ahrer}, Eva-Maria and {Alam}, Munazza K. and {Bell}, Taylor J. and {Blecic}, Jasmina and {Caceres}, Claudio and {Carter}, Aarynn L. and {Casewell}, Sarah L. and {Crouzet}, Nicolas and {Cubillos}, Patricio E. and {Decin}, Leen and {Fortney}, Jonathan J. and {Gibson}, Neale P. and {Heng}, Kevin and {Henning}, Thomas and {Iro}, Nicolas and {Kendrew}, Sarah and {Lagage}, Pierre-Olivier and {Leconte}, J{\'e}r{\'e}my and {Lendl}, Monika and {Lothringer}, Joshua D. and {Mancini}, Luigi and {Mikal-Evans}, Thomas and {Molaverdikhani}, Karan and {Nikolov}, Nikolay K. and {Ohno}, Kazumasa and {Palle}, Enric and {Piaulet}, Caroline and {Redfield}, Seth and {Roy}, Pierre-Alexis and {Tsai}, Shang-Min and {Venot}, Olivia and {Wheatley}, Peter J.},
        title = "{A broadband thermal emission spectrum of the ultra-hot Jupiter WASP-18b}",
      journal = {\nat},
         year = 2023,
        month = aug,
       volume = {620},
       number = {7973},
        pages = {292-298},
          doi = {10.1038/s41586-023-06230-1},
archivePrefix = {arXiv},
       eprint = {2301.08192},
 primaryClass = {astro-ph.EP},
       adsurl = {https://ui.adsabs.harvard.edu/abs/2023Natur.620..292C}
}

@ARTICLE{Challener2025,
       author = {{Challener}, Ryan C. and {Weiner Mansfield}, Megan and {Cubillos}, Patricio E. and {Piette}, Anjali A.~A. and {Coulombe}, Louis-Philippe and {Beltz}, Hayley and {Blecic}, Jasmina and {Rauscher}, Emily and {Bean}, Jacob L. and {Benneke}, Bj{\"o}rn and {Kempton}, Eliza M.-R. and {Harrington}, Joseph and {Komacek}, Thaddeus D. and {Parmentier}, Vivien and {Casewell}, S.~L. and {Iro}, Nicolas and {Mancini}, Luigi and {Nixon}, Matthew C. and {Radica}, Michael and {Steinrueck}, Maria E. and {Welbanks}, Luis and {Batalha}, Natalie M. and {Caceres}, Claudio and {Crossfield}, Ian J.~M. and {Crouzet}, Nicolas and {D{\'e}sert}, Jean-Michel and {Molaverdikhani}, Karan and {Nikolov}, Nikolay K. and {Palle}, Enric and {Rackham}, Benjamin V. and {Schlawin}, Everett and {Sing}, David K. and {Stevenson}, Kevin B. and {Tan}, Xianyu and {Turner}, Jake D. and {Zhang}, Xi},
        title = "{Horizontal and vertical exoplanet thermal structure from a JWST spectroscopic eclipse map}",
      journal = {arXiv e-prints},
         year = 2025,
        month = oct,
          eid = {arXiv:2510.24708},
        pages = {arXiv:2510.24708},
          doi = {10.48550/arXiv.2510.24708},
archivePrefix = {arXiv},
       eprint = {2510.24708},
 primaryClass = {astro-ph.EP},
       adsurl = {https://ui.adsabs.harvard.edu/abs/2025arXiv251024708C}
}

@ARTICLE{exoTEDRF,
       author = {{Radica}, Michael},
        title = "{exoTEDRF: An EXOplanet Transit and Eclipse Data Reduction Framework}",
      journal = {The Journal of Open Source Software},
         year = 2024,
        month = aug,
       volume = {9},
       number = {100},
          eid = {6898},
        pages = {6898},
          doi = {10.21105/joss.06898},
archivePrefix = {arXiv},
       eprint = {2407.17541},
 primaryClass = {astro-ph.IM},
       adsurl = {https://ui.adsabs.harvard.edu/abs/2024JOSS....9.6898R}
}

@ARTICLE{Radica2023,
       author = {{Radica}, Michael and {Welbanks}, Luis and {Espinoza}, N{\'e}stor and {Taylor}, Jake and {Coulombe}, Louis-Philippe and {Feinstein}, Adina D. and {Goyal}, Jayesh and {Scarsdale}, Nicholas and {Albert}, Lo{\"\i}c and {Baghel}, Priyanka and {Bean}, Jacob L. and {Blecic}, Jasmina and {Lafreni{\`e}re}, David and {MacDonald}, Ryan J. and {Zamyatina}, Maria and {Allart1}, Romain and {Artigau}, {\'E}tienne and {Batalha}, Natasha E. and {Cook}, Neil James and {Cowan}, Nicolas B. and {Dang}, Lisa and {Doyon}, Ren{\'e} and {Fournier-Tondreau}, Marylou and {Johnstone}, Doug and {Line}, Michael R. and {Moran}, Sarah E. and {Mukherjee}, Sagnick and {Pelletier}, Stefan and {Roy}, Pierre-Alexis and {Talens}, Geert Jan and {Filippazzo}, Joseph and {Pontoppidan}, Klaus and {Volk}, Kevin},
        title = "{Awesome SOSS: transmission spectroscopy of WASP-96b with NIRISS/SOSS}",
      journal = {\mnras},
         year = 2023,
        month = sep,
       volume = {524},
       number = {1},
        pages = {835-856},
          doi = {10.1093/mnras/stad1762},
archivePrefix = {arXiv},
       eprint = {2305.17001},
 primaryClass = {astro-ph.EP},
       adsurl = {https://ui.adsabs.harvard.edu/abs/2023MNRAS.524..835R}
}

@ARTICLE{Feinstein2023,
       author = {{Feinstein}, Adina D. and {Radica}, Michael and {Welbanks}, Luis and {Murray}, Catriona Anne and {Ohno}, Kazumasa and {Coulombe}, Louis-Philippe and {Espinoza}, N{\'e}stor and {Bean}, Jacob L. and {Teske}, Johanna K. and {Benneke}, Bj{\"o}rn and {Line}, Michael R. and {Rustamkulov}, Zafar and {Saba}, Arianna and {Tsiaras}, Angelos and {Barstow}, Joanna K. and {Fortney}, Jonathan J. and {Gao}, Peter and {Knutson}, Heather A. and {MacDonald}, Ryan J. and {Mikal-Evans}, Thomas and {Rackham}, Benjamin V. and {Taylor}, Jake and {Parmentier}, Vivien and {Batalha}, Natalie M. and {Berta-Thompson}, Zachory K. and {Carter}, Aarynn L. and {Changeat}, Quentin and {dos Santos}, Leonardo A. and {Gibson}, Neale P. and {Goyal}, Jayesh M. and {Kreidberg}, Laura and {L{\'o}pez-Morales}, Mercedes and {Lothringer}, Joshua D. and {Miguel}, Yamila and {Molaverdikhani}, Karan and {Moran}, Sarah E. and {Morello}, Giuseppe and {Mukherjee}, Sagnick and {Sing}, David K. and {Stevenson}, Kevin B. and {Wakeford}, Hannah R. and {Ahrer}, Eva-Maria and {Alam}, Munazza K. and {Alderson}, Lili and {Allen}, Natalie H. and {Batalha}, Natasha E. and {Bell}, Taylor J. and {Blecic}, Jasmina and {Brande}, Jonathan and {Caceres}, Claudio and {Casewell}, S.~L. and {Chubb}, Katy L. and {Crossfield}, Ian J.~M. and {Crouzet}, Nicolas and {Cubillos}, Patricio E. and {Decin}, Leen and {D{\'e}sert}, Jean-Michel and {Harrington}, Joseph and {Heng}, Kevin and {Henning}, Thomas and {Iro}, Nicolas and {Kempton}, Eliza M. -R. and {Kendrew}, Sarah and {Kirk}, James and {Krick}, Jessica and {Lagage}, Pierre-Olivier and {Lendl}, Monika and {Mancini}, Luigi and {Mansfield}, Megan and {May}, E.~M. and {Mayne}, N.~J. and {Nikolov}, Nikolay K. and {Palle}, Enric and {Petit dit de la Roche}, Dominique J.~M. and {Piaulet}, Caroline and {Powell}, Diana and {Redfield}, Seth and {Rogers}, Laura K. and {Roman}, Michael T. and {Roy}, Pierre-Alexis and {Nixon}, Matthew C. and {Schlawin}, Everett and {Tan}, Xianyu and {Tremblin}, P. and {Turner}, Jake D. and {Venot}, Olivia and {Waalkes}, William C. and {Wheatley}, Peter J. and {Zhang}, Xi},
        title = "{Early Release Science of the exoplanet WASP-39b with JWST NIRISS}",
      journal = {\nat},
         year = 2023,
        month = feb,
       volume = {614},
       number = {7949},
        pages = {670-675},
          doi = {10.1038/s41586-022-05674-1},
archivePrefix = {arXiv},
       eprint = {2211.10493},
 primaryClass = {astro-ph.EP},
       adsurl = {https://ui.adsabs.harvard.edu/abs/2023Natur.614..670F}
}

@ARTICLE{Radica2022,
       author = {{Radica}, Michael and {Albert}, Lo{\"\i}c and {Taylor}, Jake and {Lafreni{\`e}re}, David and {Coulombe}, Louis-Philippe and {Darveau-Bernier}, Antoine and {Doyon}, Ren{\'e} and {Cook}, Neil and {Cowan}, Nicolas and {Espinoza}, N{\'e}stor and {Johnstone}, Doug and {Kaltenegger}, Lisa and {Piaulet}, Caroline and {Roy}, Arpita and {Talens}, Geert Jan},
        title = "{APPLESOSS: A Producer of ProfiLEs for SOSS. Application to the NIRISS SOSS Mode}",
      journal = {\pasp},
         year = 2022,
        month = oct,
       volume = {134},
       number = {1040},
          eid = {104502},
        pages = {104502},
          doi = {10.1088/1538-3873/ac9430},
archivePrefix = {arXiv},
       eprint = {2207.05136},
 primaryClass = {astro-ph.IM},
       adsurl = {https://ui.adsabs.harvard.edu/abs/2022PASP..134j4502R}
}

@ARTICLE{Husser2013,
       author = {{Husser}, T.-O. and {Wende-von Berg}, S. and {Dreizler}, S. and {Homeier}, D. and {Reiners}, A. and {Barman}, T. and {Hauschildt}, P.~H.},
        title = "{A new extensive library of PHOENIX stellar atmospheres and synthetic spectra}",
      journal = {\aap},
         year = 2013,
        month = may,
       volume = {553},
          eid = {A6},
        pages = {A6},
          doi = {10.1051/0004-6361/201219058},
archivePrefix = {arXiv},
       eprint = {1303.5632},
 primaryClass = {astro-ph.SR},
       adsurl = {https://ui.adsabs.harvard.edu/abs/2013A&A...553A...6H}
}

@ARTICLE{Rigby2023,
       author = {{Rigby}, Jane and {Perrin}, Marshall and {McElwain}, Michael and {Kimble}, Randy and {Friedman}, Scott and {Lallo}, Matt and {Doyon}, Ren{\'e} and {Feinberg}, Lee and {Ferruit}, Pierre and {Glasse}, Alistair and {Rieke}, Marcia and {Rieke}, George and {Wright}, Gillian and {Willott}, Chris and {Colon}, Knicole and {Milam}, Stefanie and {Neff}, Susan and {Stark}, Christopher and {Valenti}, Jeff and {Abell}, Jim and {Abney}, Faith and {Abul-Huda}, Yasin and {Acton}, D. Scott and {Adams}, Evan and {Adler}, David and {Aguilar}, Jonathan and {Ahmed}, Nasif and {Albert}, Lo{\"\i}c and {Alberts}, Stacey and {Aldridge}, David and {Allen}, Marsha and {Altenburg}, Martin and {{\'A}lvarez-M{\'a}rquez}, Javier and {Alves de Oliveira}, Catarina and {Andersen}, Greg and {Anderson}, Harry and {Anderson}, Sara and {Argyriou}, Ioannis and {Armstrong}, Amber and {Arribas}, Santiago and {Artigau}, Etienne and {Arvai}, Amanda and {Atkinson}, Charles and {Bacon}, Gregory and {Bair}, Thomas and {Banks}, Kimberly and {Barrientes}, Jaclyn and {Barringer}, Bruce and {Bartosik}, Peter and {Bast}, William and {Baudoz}, Pierre and {Beatty}, Thomas and {Bechtold}, Katie and {Beck}, Tracy and {Bergeron}, Eddie and {Bergkoetter}, Matthew and {Bhatawdekar}, Rachana and {Birkmann}, Stephan and {Blazek}, Ronald and {Blome}, Claire and {Boccaletti}, Anthony and {B{\"o}ker}, Torsten and {Boia}, John and {Bonaventura}, Nina and {Bond}, Nicholas and {Bosley}, Kari and {Boucarut}, Ray and {Bourque}, Matthew and {Bouwman}, Jeroen and {Bower}, Gary and {Bowers}, Charles and {Boyer}, Martha and {Bradley}, Larry and {Brady}, Greg and {Braun}, Hannah and {Breda}, David and {Bresnahan}, Pamela and {Bright}, Stacey and {Britt}, Christopher and {Bromenschenkel}, Asa and {Brooks}, Brian and {Brooks}, Keira and {Brown}, Bob and {Brown}, Matthew and {Brown}, Patricia and {Bunker}, Andy and {Burger}, Matthew and {Bushouse}, Howard and {Cale}, Steven and {Cameron}, Alex and {Cameron}, Peter and {Canipe}, Alicia and {Caplinger}, James and {Caputo}, Francis and {Cara}, Mihai and {Carey}, Larkin and {Carniani}, Stefano and {Carrasquilla}, Maria and {Carruthers}, Margaret and {Case}, Michael and {Catherine}, Riggs and {Chance}, Don and {Chapman}, George and {Charlot}, St{\'e}phane and {Charlow}, Brian and {Chayer}, Pierre and {Chen}, Bin and {Cherinka}, Brian and {Chichester}, Sarah and {Chilton}, Zack and {Chonis}, Taylor and {Clampin}, Mark and {Clark}, Charles and {Clark}, Kerry and {Coe}, Dan and {Coleman}, Benee and {Comber}, Brian and {Comeau}, Tom and {Connolly}, Dennis and {Cooper}, James and {Cooper}, Rachel and {Coppock}, Eric and {Correnti}, Matteo and {Cossou}, Christophe and {Coulais}, Alain and {Coyle}, Laura and {Cracraft}, Misty and {Curti}, Mirko and {Cuturic}, Steven and {Davis}, Katherine and {Davis}, Michael and {Dean}, Bruce and {DeLisa}, Amy and {deMeester}, Wim and {Dencheva}, Nadia and {Dencheva}, Nadezhda and {DePasquale}, Joseph and {Deschenes}, Jeremy and {Hunor Detre}, {\"O}rs and {Diaz}, Rosa and {Dicken}, Dan and {DiFelice}, Audrey and {Dillman}, Matthew and {Dixon}, William and {Doggett}, Jesse and {Donaldson}, Tom and {Douglas}, Rob and {DuPrie}, Kimberly and {Dupuis}, Jean and {Durning}, John and {Easmin}, Nilufar and {Eck}, Weston and {Edeani}, Chinwe and {Egami}, Eiichi and {Ehrenwinkler}, Ralf and {Eisenhamer}, Jonathan and {Eisenhower}, Michael and {Elie}, Michelle and {Elliott}, James and {Elliott}, Kyle and {Ellis}, Tracy and {Engesser}, Michael and {Espinoza}, Nestor and {Etienne}, Odessa and {Etxaluze}, Mireya and {Falini}, Patrick and {Feeney}, Matthew and {Ferry}, Malcolm and {Filippazzo}, Joseph and {Fincham}, Brian and {Fix}, Mees and {Flagey}, Nicolas and {Florian}, Michael and {Flynn}, Jim and {Fontanella}, Erin and {Ford}, Terrance and {Forshay}, Peter and {Fox}, Ori and {Franz}, David and {Fu}, Henry and {Fullerton}, Alexander and {Galkin}, Sergey and {Galyer}, Anthony and {Garc{\'\i}a Mar{\'\i}n}, Macarena and {Gardner}, Jonathan P. and {Gardner}, Lisa and {Garland}, Dennis and {Garrett}, Bruce and {Gasman}, Danny and {Gaspar}, Andras and {Gaudreau}, Daniel and {Gauthier}, Peter and {Geers}, Vincent and {Geithner}, Paul and {Gennaro}, Mario and {Giardino}, Giovanna and {Girard}, Julien and {Giuliano}, Mark and {Glassmire}, Kirk and {Glauser}, Adrian},
        title = "{The Science Performance of JWST as Characterized in Commissioning}",
      journal = {\pasp},
         year = 2023,
        month = apr,
       volume = {135},
       number = {1046},
          eid = {048001},
        pages = {048001},
          doi = {10.1088/1538-3873/acb293},
archivePrefix = {arXiv},
       eprint = {2207.05632},
 primaryClass = {astro-ph.IM},
       adsurl = {https://ui.adsabs.harvard.edu/abs/2023PASP..135d8001R}
}

@ARTICLE{Brogi2023,
       author = {{Brogi}, Matteo and {Emeka-Okafor}, Vanessa and {Line}, Michael R. and {Gandhi}, Siddharth and {Pino}, Lorenzo and {Kempton}, Eliza M.-R. and {Rauscher}, Emily and {Parmentier}, Vivien and {Bean}, Jacob L. and {Mace}, Gregory N. and {Cowan}, Nicolas B. and {Shkolnik}, Evgenya and {Wardenier}, Joost P. and {Mansfield}, Megan and {Welbanks}, Luis and {Smith}, Peter and {Fortney}, Jonathan J. and {Birkby}, Jayne L. and {Zalesky}, Joseph A. and {Dang}, Lisa and {Patience}, Jennifer and {D{\'e}sert}, Jean-Michel},
        title = "{The Roasting Marshmallows Program with IGRINS on Gemini South I: Composition and Climate of the Ultrahot Jupiter WASP-18 b}",
      journal = {\aj},
         year = 2023,
        month = mar,
       volume = {165},
       number = {3},
          eid = {91},
        pages = {91},
          doi = {10.3847/1538-3881/acaf5c},
archivePrefix = {arXiv},
       eprint = {2209.15548},
 primaryClass = {astro-ph.EP},
       adsurl = {https://ui.adsabs.harvard.edu/abs/2023AJ....165...91B}
}

@ARTICLE{Yan2023,
       author = {{Yan}, F. and {Nortmann}, L. and {Reiners}, A. and {Piskunov}, N. and {Hatzes}, A. and {Seemann}, U. and {Shulyak}, D. and {Lavail}, A. and {Rains}, A.~D. and {Cont}, D. and {Rengel}, M. and {Lesjak}, F. and {Nagel}, E. and {Kochukhov}, O. and {Czesla}, S. and {Boldt-Christmas}, L. and {Heiter}, U. and {Smoker}, J.~V. and {Rodler}, F. and {Bristow}, P. and {Dorn}, R.~J. and {Jung}, Y. and {Marquart}, T. and {Stempels}, E.},
        title = "{CRIRES$^{+}$ detection of CO emissions lines and temperature inversions on the dayside of WASP-18b and WASP-76b}",
      journal = {\aap},
         year = 2023,
        month = apr,
       volume = {672},
          eid = {A107},
        pages = {A107},
          doi = {10.1051/0004-6361/202245371},
archivePrefix = {arXiv},
       eprint = {2302.08736},
 primaryClass = {astro-ph.EP},
       adsurl = {https://ui.adsabs.harvard.edu/abs/2023A&A...672A.107Y}
}

@ARTICLE{Maxted2013,
       author = {{Maxted}, P.~F.~L. and {Anderson}, D.~R. and {Doyle}, A.~P. and {Gillon}, M. and {Harrington}, J. and {Iro}, N. and {Jehin}, E. and {Lafreni{\`e}re}, D. and {Smalley}, B. and {Southworth}, J.},
        title = "{Spitzer 3.6 and 4.5 {\ensuremath{\mu}}m full-orbit light curves of WASP-18}",
      journal = {\mnras},
         year = 2013,
        month = jan,
       volume = {428},
       number = {3},
        pages = {2645-2660},
          doi = {10.1093/mnras/sts231},
archivePrefix = {arXiv},
       eprint = {1210.5585},
 primaryClass = {astro-ph.EP},
       adsurl = {https://ui.adsabs.harvard.edu/abs/2013MNRAS.428.2645M}
}

@ARTICLE{Molliere2019,
       author = {{Molli{\`e}re}, P. and {Wardenier}, J.~P. and {van Boekel}, R. and {Henning}, Th. and {Molaverdikhani}, K. and {Snellen}, I.~A.~G.},
        title = "{petitRADTRANS. A Python radiative transfer package for exoplanet characterization and retrieval}",
      journal = {\aap},
         year = 2019,
        month = jul,
       volume = {627},
          eid = {A67},
        pages = {A67},
          doi = {10.1051/0004-6361/201935470},
archivePrefix = {arXiv},
       eprint = {1904.11504},
 primaryClass = {astro-ph.EP},
       adsurl = {https://ui.adsabs.harvard.edu/abs/2019A&A...627A..67M}
}

@ARTICLE{Blain2024,
       author = {{Blain}, Doriann and {Molli{\`e}re}, Paul and {Nasedkin}, Evert},
        title = "{SpectralModel: a high-resolution framework for petitRADTRANS 3}",
      journal = {The Journal of Open Source Software},
         year = 2024,
        month = oct,
       volume = {9},
       number = {102},
          eid = {7028},
        pages = {7028},
          doi = {10.21105/joss.07028},
       adsurl = {https://ui.adsabs.harvard.edu/abs/2024JOSS....9.7028B}
}

@ARTICLE{Cortes-Zuleta2020,
       author = {{Cort{\'e}s-Zuleta}, P{\'\i}a and {Rojo}, Patricio and {Wang}, Songhu and {Hinse}, Tobias C. and {Hoyer}, Sergio and {Sanhueza}, Bastian and {Correa-Amaro}, Patricio and {Albornoz}, Julio},
        title = "{TraMoS. V. Updated ephemeris and multi-epoch monitoring of the hot Jupiters WASP-18Ab, WASP-19b, and WASP-77Ab}",
      journal = {\aap},
         year = 2020,
        month = apr,
       volume = {636},
          eid = {A98},
        pages = {A98},
          doi = {10.1051/0004-6361/201936279},
archivePrefix = {arXiv},
       eprint = {2001.11112},
 primaryClass = {astro-ph.EP},
       adsurl = {https://ui.adsabs.harvard.edu/abs/2020A&A...636A..98C}
}

@ARTICLE{Madhusudhan2009,
       author = {{Madhusudhan}, N. and {Seager}, S.},
        title = "{A Temperature and Abundance Retrieval Method for Exoplanet Atmospheres}",
      journal = {\apj},
         year = 2009,
        month = dec,
       volume = {707},
       number = {1},
        pages = {24-39},
          doi = {10.1088/0004-637X/707/1/24},
archivePrefix = {arXiv},
       eprint = {0910.1347},
 primaryClass = {astro-ph.EP},
       adsurl = {https://ui.adsabs.harvard.edu/abs/2009ApJ...707...24M}
}

@ARTICLE{Rothman2010,
       author = {{Rothman}, L.~S. and {Gordon}, I.~E. and {Barber}, R.~J. and {Dothe}, H. and {Gamache}, R.~R. and {Goldman}, A. and {Perevalov}, V.~I. and {Tashkun}, S.~A. and {Tennyson}, J.},
        title = "{HITEMP, the high-temperature molecular spectroscopic database}",
      journal = {\jqsrt},
         year = 2010,
        month = oct,
       volume = {111},
        pages = {2139-2150},
          doi = {10.1016/j.jqsrt.2010.05.001},
       adsurl = {https://ui.adsabs.harvard.edu/abs/2010JQSRT.111.2139R}
}

@ARTICLE{Yurchenko2014,
       author = {{Yurchenko}, Sergei N. and {Tennyson}, Jonathan},
        title = "{ExoMol line lists - IV. The rotation-vibration spectrum of methane up to 1500 K}",
      journal = {\mnras},
         year = 2014,
        month = may,
       volume = {440},
       number = {2},
        pages = {1649-1661},
          doi = {10.1093/mnras/stu326},
archivePrefix = {arXiv},
       eprint = {1401.4852},
 primaryClass = {astro-ph.EP},
       adsurl = {https://ui.adsabs.harvard.edu/abs/2014MNRAS.440.1649Y}
}

@ARTICLE{Rothman2013,
       author = {{Rothman}, L.~S. and {Gordon}, I.~E. and {Babikov}, Y. and {Barbe}, A. and {Chris Benner}, D. and {Bernath}, P.~F. and {Birk}, M. and {Bizzocchi}, L. and {Boudon}, V. and {Brown}, L.~R. and {Campargue}, A. and {Chance}, K. and {Cohen}, E.~A. and {Coudert}, L.~H. and {Devi}, V.~M. and {Drouin}, B.~J. and {Fayt}, A. and {Flaud}, J.-M. and {Gamache}, R.~R. and {Harrison}, J.~J. and {Hartmann}, J.-M. and {Hill}, C. and {Hodges}, J.~T. and {Jacquemart}, D. and {Jolly}, A. and {Lamouroux}, J. and {Le Roy}, R.~J. and {Li}, G. and {Long}, D.~A. and {Lyulin}, O.~M. and {Mackie}, C.~J. and {Massie}, S.~T. and {Mikhailenko}, S. and {M{\"u}ller}, H.~S.~P. and {Naumenko}, O.~V. and {Nikitin}, A.~V. and {Orphal}, J. and {Perevalov}, V. and {Perrin}, A. and {Polovtseva}, E.~R. and {Richard}, C. and {Smith}, M.~A.~H. and {Starikova}, E. and {Sung}, K. and {Tashkun}, S. and {Tennyson}, J. and {Toon}, G.~C. and {Tyuterev}, Vl. G. and {Wagner}, G.},
        title = "{The HITRAN2012 molecular spectroscopic database}",
      journal = {\jqsrt},
         year = 2013,
        month = nov,
       volume = {130},
        pages = {4-50},
          doi = {10.1016/j.jqsrt.2013.07.002},
       adsurl = {https://ui.adsabs.harvard.edu/abs/2013JQSRT.130....4R}
}

@ARTICLE{Harris2006,
       author = {{Harris}, G.~J. and {Tennyson}, J. and {Kaminsky}, B.~M. and {Pavlenko}, Ya. V. and {Jones}, H.~R.~A.},
        title = "{Improved HCN/HNC linelist, model atmospheres and synthetic spectra for WZ Cas}",
      journal = {\mnras},
         year = 2006,
        month = mar,
       volume = {367},
       number = {1},
        pages = {400-406},
          doi = {10.1111/j.1365-2966.2005.09960.x},
archivePrefix = {arXiv},
       eprint = {astro-ph/0512363},
 primaryClass = {astro-ph},
       adsurl = {https://ui.adsabs.harvard.edu/abs/2006MNRAS.367..400H}
}

@ARTICLE{Bernath2020,
       author = {{Bernath}, Peter F.},
        title = "{MoLLIST: Molecular Line Lists, Intensities and Spectra}",
      journal = {\jqsrt},
         year = 2020,
        month = jan,
       volume = {240},
          eid = {106687},
        pages = {106687},
          doi = {10.1016/j.jqsrt.2019.106687},
       adsurl = {https://ui.adsabs.harvard.edu/abs/2020JQSRT.24006687B}
}

@ARTICLE{Polyansky2018,
       author = {{Polyansky}, Oleg L. and {Kyuberis}, Aleksandra A. and {Zobov}, Nikolai F. and {Tennyson}, Jonathan and {Yurchenko}, Sergei N. and {Lodi}, Lorenzo},
        title = "{ExoMol molecular line lists XXX: a complete high-accuracy line list for water}",
      journal = {\mnras},
         year = 2018,
        month = oct,
       volume = {480},
       number = {2},
        pages = {2597-2608},
          doi = {10.1093/mnras/sty1877},
archivePrefix = {arXiv},
       eprint = {1807.04529},
 primaryClass = {astro-ph.EP},
       adsurl = {https://ui.adsabs.harvard.edu/abs/2018MNRAS.480.2597P}
}

@INPROCEEDINGS{Kurucz2018,
       author = {{Kurucz}, R.~L.},
        title = "{Including All the Lines: Data Releases for Spectra and Opacities through 2017}",
    booktitle = {Workshop on Astrophysical Opacities},
         year = 2018,
       series = {Astronomical Society of the Pacific Conference Series},
       volume = {515},
        month = aug,
        pages = {47},
       adsurl = {https://ui.adsabs.harvard.edu/abs/2018ASPC..515...47K}
}

@BOOK{Gray2008,
       author = {{Gray}, David F.},
        title = "{The Observation and Analysis of Stellar Photospheres}",
         year = 2008,
       adsurl = {https://ui.adsabs.harvard.edu/abs/2008oasp.book.....G}
}

@ARTICLE{Kitzmann2024,
       author = {{Kitzmann}, Daniel and {Stock}, Joachim W. and {Patzer}, A. Beate C.},
        title = "{FASTCHEM COND: equilibrium chemistry with condensation and rainout for cool planetary and stellar environments}",
      journal = {\mnras},
         year = 2024,
        month = jan,
       volume = {527},
       number = {3},
        pages = {7263-7283},
          doi = {10.1093/mnras/stad3515},
archivePrefix = {arXiv},
       eprint = {2309.02337},
 primaryClass = {astro-ph.EP},
       adsurl = {https://ui.adsabs.harvard.edu/abs/2024MNRAS.527.7263K}
}

@INPROCEEDINGS{Skilling2004,
       author = {{Skilling}, John},
        title = "{Nested Sampling}",
    booktitle = {Bayesian Inference and Maximum Entropy Methods in Science and Engineering: 24th International Workshop on Bayesian Inference and Maximum Entropy Methods in Science and Engineering},
         year = 2004,
       editor = {{Fischer}, Rainer and {Preuss}, Roland and {Toussaint}, Udo Von},
       series = {American Institute of Physics Conference Series},
       volume = {735},
        month = nov,
    publisher = {AIP},
        pages = {395-405},
          doi = {10.1063/1.1835238},
       adsurl = {https://ui.adsabs.harvard.edu/abs/2004AIPC..735..395S}
}

@ARTICLE{Speagle2020,
       author = {{Speagle}, Joshua S.},
        title = "{DYNESTY: a dynamic nested sampling package for estimating Bayesian posteriors and evidences}",
      journal = {\mnras},
         year = 2020,
        month = apr,
       volume = {493},
       number = {3},
        pages = {3132-3158},
          doi = {10.1093/mnras/staa278},
archivePrefix = {arXiv},
       eprint = {1904.02180},
 primaryClass = {astro-ph.IM},
       adsurl = {https://ui.adsabs.harvard.edu/abs/2020MNRAS.493.3132S}
}

@ARTICLE{Yan2022,
       author = {{Yan}, F. and {Pall{\'e}}, E. and {Reiners}, A. and {Casasayas-Barris}, N. and {Cont}, D. and {Stangret}, M. and {Nortmann}, L. and {Molli{\`e}re}, P. and {Henning}, Th. and {Chen}, G. and {Molaverdikhani}, K.},
        title = "{Detection of CO emission lines in the dayside atmospheres of WASP-33b and WASP-189b with GIANO}",
      journal = {\aap},
         year = 2022,
        month = may,
       volume = {661},
          eid = {L6},
        pages = {L6},
          doi = {10.1051/0004-6361/202243503},
archivePrefix = {arXiv},
       eprint = {2204.10158},
 primaryClass = {astro-ph.EP},
       adsurl = {https://ui.adsabs.harvard.edu/abs/2022A&A...661L...6Y}
}

@ARTICLE{Cont2022,
       author = {{Cont}, D. and {Yan}, F. and {Reiners}, A. and {Nortmann}, L. and {Molaverdikhani}, K. and {Pall{\'e}}, E. and {Henning}, Th. and {Ribas}, I. and {Quirrenbach}, A. and {Caballero}, J.~A. and {Amado}, P.~J. and {Czesla}, S. and {Lesjak}, F. and {L{\'o}pez-Puertas}, M. and {Molli{\`e}re}, P. and {Montes}, D. and {Morello}, G. and {Nagel}, E. and {Pedraz}, S. and {S{\'a}nchez-L{\'o}pez}, A.},
        title = "{Atmospheric characterization of the ultra-hot Jupiter WASP-33b. Detection of Ti and V emission lines and retrieval of a broadened line profile}",
      journal = {\aap},
         year = 2022,
        month = dec,
       volume = {668},
          eid = {A53},
        pages = {A53},
          doi = {10.1051/0004-6361/202244277},
archivePrefix = {arXiv},
       eprint = {2209.10618},
 primaryClass = {astro-ph.EP},
       adsurl = {https://ui.adsabs.harvard.edu/abs/2022A&A...668A..53C}
}

@ARTICLE{Parmentier2018,
       author = {{Parmentier}, Vivien and {Line}, Mike R. and {Bean}, Jacob L. and {Mansfield}, Megan and {Kreidberg}, Laura and {Lupu}, Roxana and {Visscher}, Channon and {D{\'e}sert}, Jean-Michel and {Fortney}, Jonathan J. and {Deleuil}, Magalie and {Arcangeli}, Jacob and {Showman}, Adam P. and {Marley}, Mark S.},
        title = "{From thermal dissociation to condensation in the atmospheres of ultra hot Jupiters: WASP-121b in context}",
      journal = {\aap},
         year = 2018,
        month = sep,
       volume = {617},
          eid = {A110},
        pages = {A110},
          doi = {10.1051/0004-6361/201833059},
archivePrefix = {arXiv},
       eprint = {1805.00096},
 primaryClass = {astro-ph.EP},
       adsurl = {https://ui.adsabs.harvard.edu/abs/2018A&A...617A.110P}
}

@ARTICLE{Polanski2022,
       author = {{Polanski}, Alex S. and {Crossfield}, Ian J.~M. and {Howard}, Andrew W. and {Isaacson}, Howard and {Rice}, Malena},
        title = "{Chemical Abundances for 25 JWST Exoplanet Host Stars with KeckSpec}",
      journal = {Research Notes of the American Astronomical Society},
         year = 2022,
        month = aug,
       volume = {6},
       number = {8},
          eid = {155},
        pages = {155},
          doi = {10.3847/2515-5172/ac8676},
archivePrefix = {arXiv},
       eprint = {2207.13662},
 primaryClass = {astro-ph.EP},
       adsurl = {https://ui.adsabs.harvard.edu/abs/2022RNAAS...6..155P}
}

@ARTICLE{Oberg2011,
       author = {{{\"O}berg}, Karin I. and {Murray-Clay}, Ruth and {Bergin}, Edwin A.},
        title = "{The Effects of Snowlines on C/O in Planetary Atmospheres}",
      journal = {\apjl},
         year = 2011,
        month = dec,
       volume = {743},
       number = {1},
          eid = {L16},
        pages = {L16},
          doi = {10.1088/2041-8205/743/1/L16},
archivePrefix = {arXiv},
       eprint = {1110.5567},
 primaryClass = {astro-ph.GA},
       adsurl = {https://ui.adsabs.harvard.edu/abs/2011ApJ...743L..16O}
}

@software{Czesla2019,
       author = {{Czesla}, Stefan and {Schr{\"o}ter}, Sebastian and {Schneider}, Christian P. and {Huber}, Klaus F. and {Pfeifer}, Fabian and {Andreasen}, Daniel T. and {Zechmeister}, Mathias},
        title = "{PyA: Python astronomy-related packages}",
 howpublished = {Astrophysics Source Code Library, record ascl:1906.010},
         year = 2019,
        month = jun,
          eid = {ascl:1906.010},
archivePrefix = {ascl},
       eprint = {1906.010},
       adsurl = {https://ui.adsabs.harvard.edu/abs/2019ascl.soft06010C}
}

@ARTICLE{Virtanen2020,
       author = {{Virtanen}, Pauli and {Gommers}, Ralf and {Oliphant}, Travis E. and {Haberland}, Matt and {Reddy}, Tyler and {Cournapeau}, David and {Burovski}, Evgeni and {Peterson}, Pearu and {Weckesser}, Warren and {Bright}, Jonathan and {van der Walt}, St{\'e}fan J. and {Brett}, Matthew and {Wilson}, Joshua and {Millman}, K. Jarrod and {Mayorov}, Nikolay and {Nelson}, Andrew R.~J. and {Jones}, Eric and {Kern}, Robert and {Larson}, Eric and {Carey}, C.~J. and {Polat}, {\.I}lhan and {Feng}, Yu and {Moore}, Eric W. and {VanderPlas}, Jake and {Laxalde}, Denis and {Perktold}, Josef and {Cimrman}, Robert and {Henriksen}, Ian and {Quintero}, E.~A. and {Harris}, Charles R. and {Archibald}, Anne M. and {Ribeiro}, Ant{\^o}nio H. and {Pedregosa}, Fabian and {van Mulbregt}, Paul and {SciPy 1. 0 Contributors}},
        title = "{SciPy 1.0: fundamental algorithms for scientific computing in Python}",
      journal = {Nature Medicine},
         year = 2020,
        month = feb,
       volume = {17},
        pages = {261-272},
          doi = {10.1038/s41592-019-0686-2},
archivePrefix = {arXiv},
       eprint = {1907.10121},
 primaryClass = {cs.MS},
       adsurl = {https://ui.adsabs.harvard.edu/abs/2020NatMe..17..261V}
}

@ARTICLE{Astropy_Collaboration2013,
       author = {{Astropy Collaboration} and {Robitaille}, Thomas P. and {Tollerud}, Erik J. and {Greenfield}, Perry and {Droettboom}, Michael and {Bray}, Erik and {Aldcroft}, Tom and {Davis}, Matt and {Ginsburg}, Adam and {Price-Whelan}, Adrian M. and {Kerzendorf}, Wolfgang E. and {Conley}, Alexander and {Crighton}, Neil and {Barbary}, Kyle and {Muna}, Demitri and {Ferguson}, Henry and {Grollier}, Fr{\'e}d{\'e}ric and {Parikh}, Madhura M. and {Nair}, Prasanth H. and {Unther}, Hans M. and {Deil}, Christoph and {Woillez}, Julien and {Conseil}, Simon and {Kramer}, Roban and {Turner}, James E.~H. and {Singer}, Leo and {Fox}, Ryan and {Weaver}, Benjamin A. and {Zabalza}, Victor and {Edwards}, Zachary I. and {Azalee Bostroem}, K. and {Burke}, D.~J. and {Casey}, Andrew R. and {Crawford}, Steven M. and {Dencheva}, Nadia and {Ely}, Justin and {Jenness}, Tim and {Labrie}, Kathleen and {Lim}, Pey Lian and {Pierfederici}, Francesco and {Pontzen}, Andrew and {Ptak}, Andy and {Refsdal}, Brian and {Servillat}, Mathieu and {Streicher}, Ole},
        title = "{Astropy: A community Python package for astronomy}",
      journal = {\aap},
         year = 2013,
        month = oct,
       volume = {558},
          eid = {A33},
        pages = {A33},
          doi = {10.1051/0004-6361/201322068},
archivePrefix = {arXiv},
       eprint = {1307.6212},
 primaryClass = {astro-ph.IM},
       adsurl = {https://ui.adsabs.harvard.edu/abs/2013A&A...558A..33A}
}
\bibliographystyle{aasjournalv7}



\end{document}